\documentclass{aa}  

\usepackage{graphicx}
\usepackage{txfonts}
\usepackage{amsmath}
\usepackage{lscape}
\usepackage{longtable}
\usepackage{amssymb}	
\usepackage{rotating}
\usepackage{lineno}
\usepackage{natbib}
\usepackage[colorlinks,linkcolor=blue,anchorcolor=blue,citecolor=blue]{hyperref}

\newcommand{\fcont}{{$f_\mathrm{cont}$}}

\newcommand\T{\rule{0pt}{2.6ex}} 
\newcommand\B{\rule[-1.2ex]{0pt}{0pt}} 
\begin{document}

   \title{The ACT-DR5~MCMF Galaxy Cluster Catalog}


   \author{M. Klein
          \inst{1},
          J.J. Mohr\inst{1,}\inst{2}
          \and
          C.T. Davies\inst{1} 
          }

   \institute{University Observatory, Faculty of Physics, Ludwig-Maximilians-Universit{\"a}t, Scheinerstr. 1, 81679, Munich, Germany\\
              \email{matthias.klein@physik.lmu.de}
         \and         
            Max Planck Institute for Extraterrestrial Physics, Giessenbachstrasse 1, 85748 Garching, Germany 
            }

   \date{xxx; accepted xx}

\titlerunning{ACT-DR5~MCMF Clusters}
\authorrunning{Klein et al.}

 
  \abstract
{Galaxy clusters are useful cosmological probes and interesting astrophysical laboratories.  As the cluster samples continue to grow in size, a deeper understanding of the sample characteristics and improved control of systematics becomes more crucial. 
In this analysis we create a new and larger ACT-DR5-based thermal Sunyaev–Zel’dovich Effect (tSZE) selected galaxy cluster catalog with improved control over sample purity and completeness. We employ the red sequence based cluster redshift and confirmation tool MCMF together with optical imaging data from the Legacy Survey DR-10 and infrared data from the WISE satellite to systematically identify true clusters from a new cluster candidate detection run on the ACT-DR5 dataset.
The resulting ACT-DR5~MCMF sample contains 6,237 clusters with a residual contamination of 10.7\%. This is an increase of 51\% compared to the previous ACT-DR5 cluster catalog, making this new catalog the largest tSZE-selected cluster catalog to date.  The $z_\mathrm{phot}$>1 subsample contains 703 clusters, three times more than in the previous ACT-DR5 catalog. Cross-matching the ACT-DR5~MCMF cluster catalog with a deeper tSZE sample from SPTpol~500d allows us to confirm the completeness and purity of the new ACT-DR5~MCMF sample.  Cross-matching to the two largest X-ray selected cluster samples, the all-sky RASS~MCMF and the western Galactic hemisphere survey eRASS1, confirms the sample purity of the RASS~MCMF sample and in the case of eROSITA eRASS1 reveals that 43\% of the matched clusters are designated in eRASS1 as X-ray point sources rather than groups and clusters. Cross-correlating the ACT-DR5~MCMF cluster catalog with ACT-DR6 lensing maps results in a 16.4\,$\sigma$ detection of CMB lensing around the clusters, corresponding to the strongest signal found so far for a galaxy cluster sample. Repeating the measurement for the $z>1$ cluster subsample yields a significance of 4.3\,$\sigma$, which is the strongest CMB lensing detection in a z>1 cluster sample to date.}


   \maketitle
%
\section{Introduction}
The thermal Sunyaev–Zel’dovich Effect \citep[tSZE; ][]{Sunyaev72} has long been recognized as a means of creating nearly mass-limited samples of galaxy clusters, independent of their redshift \citep{Barbosa98,Haiman01}. The first tSZE-selected clusters were reported more than a decade ago \citep{Staniszewski2009ApJ...701...32S}, and in recent years, there has been significant progress in utilizing tSZE-selected galaxy clusters together with weak gravitational lensing datasets to study cosmic acceleration, to probe the so-called S8 tension \citep[e.g.,][]{PlanckSZcosmology,Bocquet2019ApJ...878...55B,Planck2020LegacyCosmology,Bocquet2024arXiv240102075B} and to explore the promise of tSZE selected cluster samples for constraining modified gravity and interacting dark matter models \citep[e.g.,][]{Vogt2024PhRvD.109l3503V,Mazoun2024PhRvD.109f3536M}.

The tSZE signature-- as well as the X-ray emission from galaxy clusters-- arises from the hot plasma or intracluster medium (ICM) that compromises $\sim$15\% of the cluster mass. The tSZE signal depends on the electron pressure of the ICM and therefore preferentially traces massive collapsed structures like the centers of clusters. This is in contrast to selecting galaxy clusters through overdensities of passive galaxies \citep{gladders00,Rykoff14}, which is sensitive to collections of passive galaxies whether they lie within a cluster halo or in the surrounding structures far outside the cluster.  
This lack of contrast in galaxy-based cluster selection complicates the cosmological modeling of such samples. Ways to account for this challenge have been developed with mixed success \citep[e.g.,][]{costanzi19, CAMIRA-WL, DESY1clucosmo, KidsClusterCosmo}, and this challenge remains as one of the key issues for the success of future optically selected cluster analyses using data from the Rubin LSST \citep{LSSToverview} and Euclid surveys \citep{EuclidMission}. 

Of the two leading ICM-based cluster selection methods, the tSZE offers some advantages. First, the tSZE signature of a cluster is strongly dependent on cluster mass and approximately independent of cluster redshift, and so the selected cluster sample has an effective mass selection that is only weakly dependent on redshift for $z>0.2$. Second, 
the tSZE signature appears as a shadow (negative signal) at frequencies below 200\,GHz. Due to the uniqueness of this signature, catalogs of tSZE selected cluster candidates are solely contaminated by noise fluctuations in the mm-wave maps. This has the advantage that the fraction of contaminants as a function of tSZE detection signal to noise ratio (SNR) can be accurately predicted, and it falls rapidly with increasing tSZE SNR. Combined with the redshift independence of the tSZE, this has the benefit that no noise fluctuations can be incorrectly interpreted as massive clusters. 

This is in contrast to X-ray selected cluster samples, where other physical X-ray sources (AGN, Stars) exhibit similar X-ray fluxes and are significantly more frequent than clusters. Due to the strong redshift dependence of the X-ray signature of a cluster, even a low significance source can be mistakenly interpreted to be a high mass cluster if it is at high redshift. Imposing a selection on source extent to exclude point sources like AGN and stars helps to increase the purity of the candidate list if the X-ray detector has sufficient angular resolution, but it comes at the price of a significant reduction of the number of selected clusters and a more complicated selection function. Also, even with X-ray extent selection the sample still includes significant contamination that requires modeling or cleaning  \citep{Klein22,Chiu23}. The challenges associated with X-ray selection become more important with larger cluster samples, as recently highlighted in the eROSITA eRASS1 cosmological study \citep{eRASS1cosmo}.

In this work, we are building upon our previous body of work creating large ICM-selected cluster catalogs for the RASS, SPT, Planck and eROSITA eFEDS surveys. \citep{Klein19,Klein22,Hernandez22,Klein23,Klein24a,Bleem24}. Motivated by the successful cluster-based cosmological analyses \citep{Chiu23,Bocquet2024arXiv240102075B} of the two most recent catalogs, we apply our Multi-Component Matched Filter (MCMF) cluster confirmation technique \citep{Klein18} to tSZE data from the Atacama Cosmology Telescope (ACT) data release 5 \citep[ACT-DR5,][]{ACTDR5maps}. This dataset covers 13,211\,deg$^2$ of extra galactic sky and has been previously used to create a cluster catalog of 4,195 tSZE-selected clusters \citep{ACTDR5}. Our main motivation in reanalysing this sample is to create a cluster catalog with more homogeneous and complete optical confirmation so that we can quantify the impact on the completeness and contamination of the final cluster sample and estimate the level of contamination in the tSZE selected candidate list. Thus, we expect to add value to the cluster catalog, enabling and ideally simplifying cosmological studies employing this sample.

Throughout this paper we assume a flat $\Lambda$CDM model with $\Omega_\mathrm{m}=0.3$ and $H_0=70$~km$\,$s$^{-1}\,$Mpc$^{-1}$ unless otherwise noted.

\begin{figure}
\begin{center}
\includegraphics[width=0.99\linewidth]{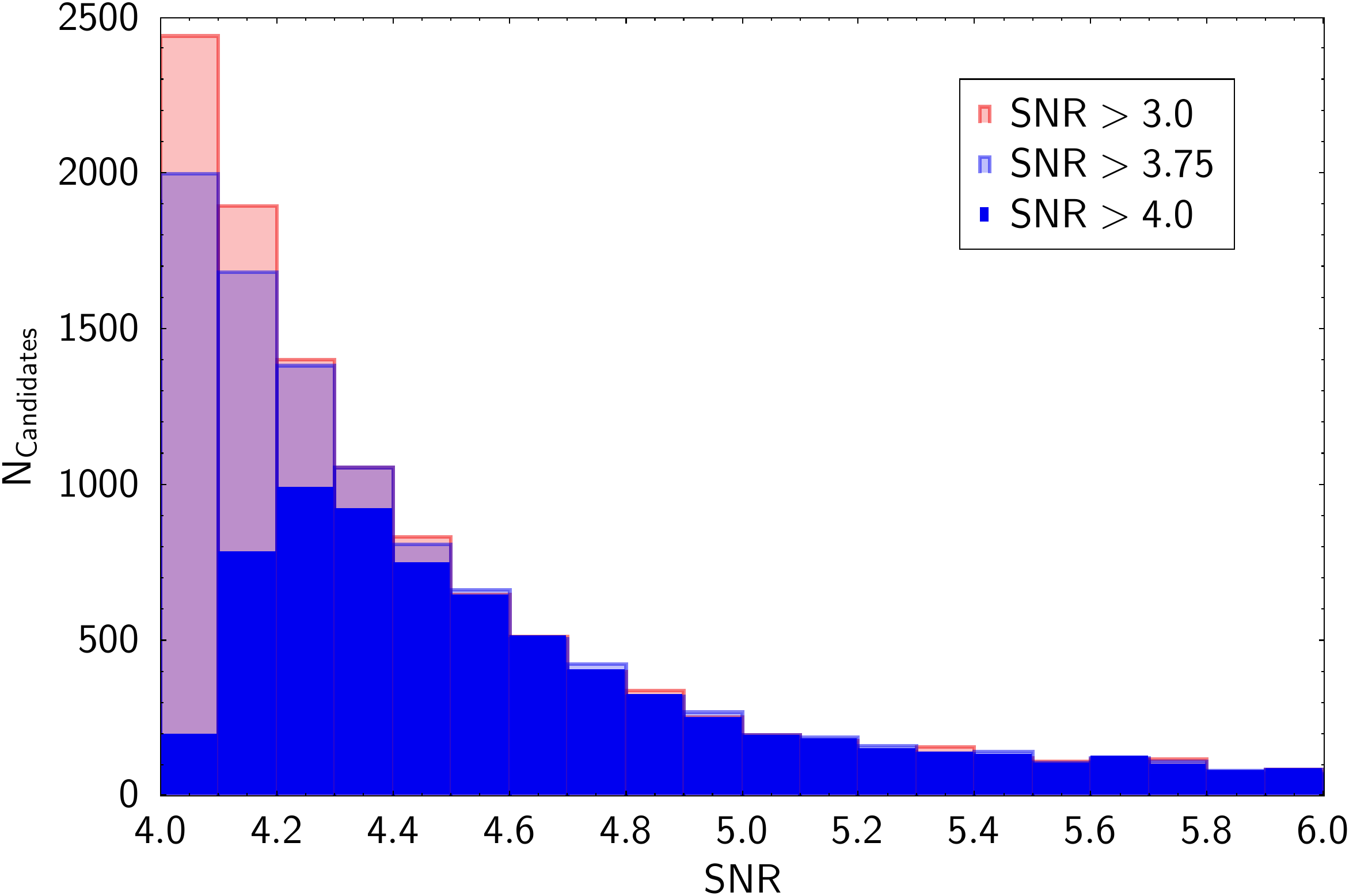}
\caption{Distribution of ACT cluster candidates in tSZE detection signal to noise ratio (SNR) for different source detection thresholds. The threshold SNR>4 was used in the previous analysis \citep{ACTDR5}, while here we adopt SNR>3 for detection.}
\label{fig:ACTsnrdet}
\end{center}
\end{figure}

\begin{figure*}
\begin{center}
\includegraphics[width=0.99\linewidth]{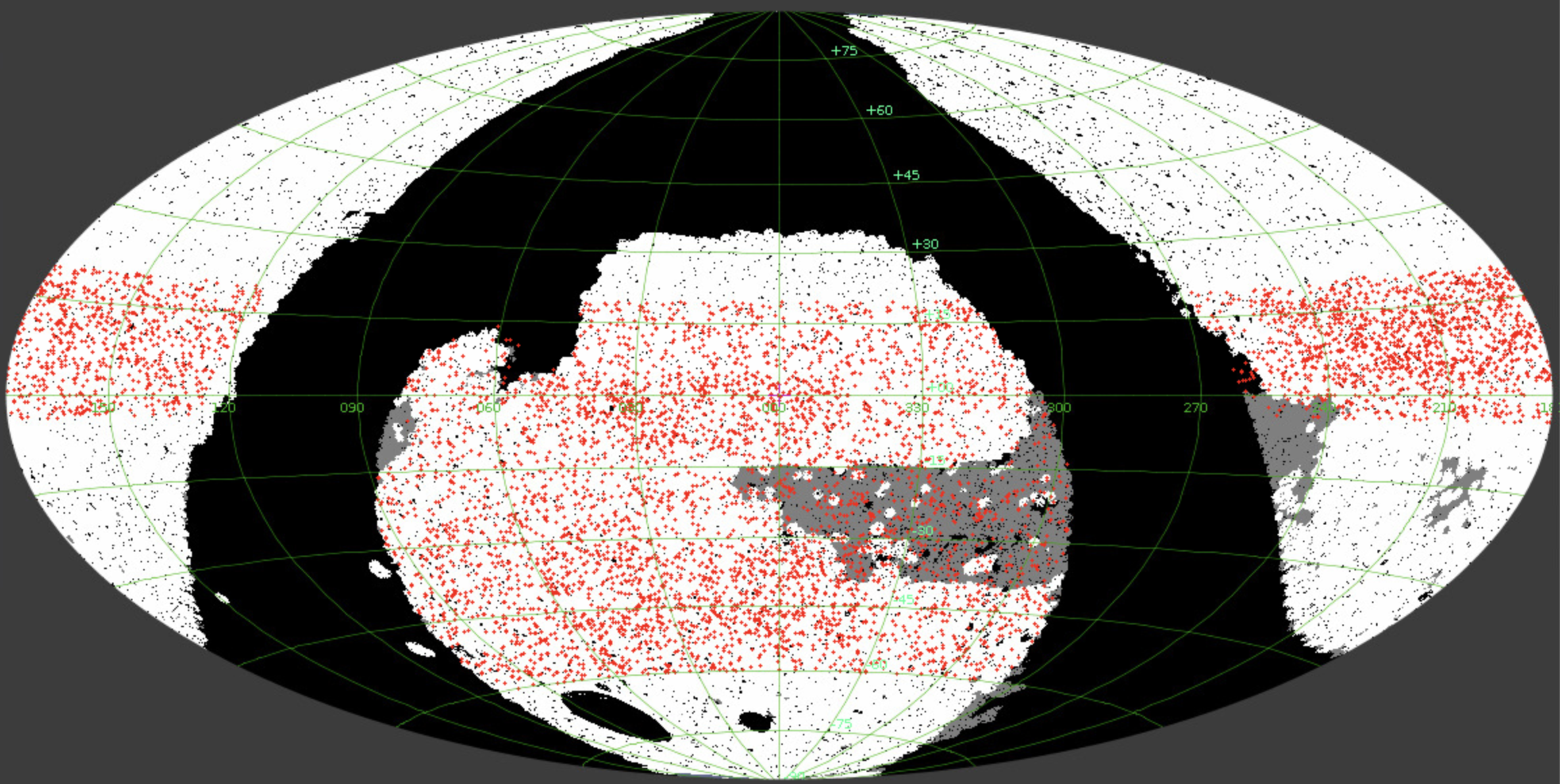}
\caption{Optical footprint of the LS-DR10 dataset and distribution of ACT-DR5 sources (red points). White regions have $g,r,z,w1,w2$-band coverage, while grey regions have $g,i,w1,w2$-band coverage.}\label{fig:ACTfoot}
\end{center}
\end{figure*}
\section{Datasets}

\subsection{The ACT-DR5 dataset}\label{sec:ACTdata}
The dataset, footprint and detection algorithm used to create the list of tSZE candidates is almost identical to those used to create the previously published list of cluster candidates \citep{ACTDR5}. We therefore restrict this section to a high-level overview and focus on the differences between the \citet{ACTDR5} analysis and this work. For more details on the tSZE datasets we refer the interested reader to \cite{ACTDR5} and references therein.

Similar to \cite{ACTDR5} we make use of the fifth data release of ACT data, which contains data obtained between 2008 and 2018. It includes data obtained using the AdvACT camera \citep[][]{AdvAct} starting in 2016. For the tSZE source detection we use the 98 and 150\,GHz channels, which have approximate beam sizes of 2.2 and 1.4\,arcmin FWHM. The coadded maps cover $\approx 18,000$\, deg$^2$, but following \cite{ACTDR5} we exclude bad regions such as regions of Galactic latitudes ($| b | < 20$\,deg), regions around point sources and dusty regions. The remaining useful area 
is 13,211\,deg$^2$, from which $\approx 12,000$\,deg$^2$ have noise levels of $<30\ \mu\mathrm{K}$.

For the source detection we use the NEMO\footnote{see https://github.com/simonsobs/nemo/}
pipeline \citep{ACTDR5}.  The NEMO algorithm includes a multi-frequency matched filter \citep{Melin06, Williamson11} applied to the 98 and 150\,GHz maps that includes a cluster based spatial and spectral filter along with the noise covariance between maps.
The spectral filter is based upon the cluster tSZE spectrum and the spatial filter is based on the universal pressure profile \citep{Arnaud10} convolved with the beam appropriate for each frequency. To account for different cluster sizes, 16 different matched filters are applied, corresponding to cluster models with masses $\mathrm{M}_\mathrm{500c}$ of 1, 2, 4 and 8 $\times 10^{14}\ \mathrm{M}_{\odot}$ at four different redshifts (0.2, 0.4, 0.8, 1.2).   

The version of NEMO used to construct our cluster candidate list differs slightly from that used in the original 2021 analysis \citep{ACTDR5}. We first run the NEMO code with the same settings used in \cite{ACTDR5},  which yields 8,727 candidates, which differs by only 1.7\% from that reported in \cite{ACTDR5} (8,878). This gives us confidence that any changes in the pipeline have only a minor impact on the candidate list.
Examining the distribution of candidates in the reported signal to noise ratio (SNR) (see Fig.~\ref{fig:ACTsnrdet}), we find that the distribution drops at SNR<4.3, well above the nominal detection threshold of SNR=4.  Because the expected number of noise fluctuations as well as that of real sources should both continuously grow towards lower SNR values, this behaviour suggests an enhanced incompleteness in the candidate list when approaching the specified detection threshold. 
To avoid this effect, we 
run the source detection with lower detection thresholds. We find that the source list is stable for sources with SNR that lie $\sim$0.5 above the detection threshold. 
To guarantee a truly SNR selected sample, we define our candidate list as all sources above SNR>4 from a source detection run that uses a threshold of SNR=3. In contrast to the original threshold this setting provides a list of 12,671 candidates, 45\% more than are found when running with a threshold of SNR=4.

\subsection{Legacy Survey (LS)}

Similar to our previous work on ROSAT X-ray selected galaxy clusters \citep{Klein23}, we use the most recent data release (DR10) from the DESI Legacy Imaging Surveys \citep[LS;][]{Legacysurveys19}.
Up to data release 8, LS was a combination of four imaging surveys, the 9,000 $\text{deg}^2$ $grz$-band DECam \citep{Flaugher15} based DECaLS survey, the 5,000 $\text{deg}^2$ BASS and MzLS surveys providing photometry in $g,r$ and $z$-band, respectively, and the WISE and NEOWISE surveys in the mid-IR at 3.4\,$\mu$m and 4.6\,$\mu$m. 
LS-DR10 includes all imaging data from the Dark Energy Survey as well as from various other survey programs such as BLISS and the DeROSITAS surveys, which resulted in a significant increase of the total survey area compared to DR9.
Compared to previous data releases DR10 also includes DECam imaging data from the $i$-band, increasing the amount of color information of sources over significant parts of the survey with declination below 30\,deg.
In Fig.~\ref{fig:ACTfoot} we show the LS-DR10 footprint along with the distribution of ACT-DR5 sources in red.

The imaging depth depends on sky position. The 5,000~deg$^2$ BASS and MzLS surveys, covering the sky at Declination >30\,deg (Legacy Survey North), exhibits a typical 5$\sigma$ point source depth of $\sim$24.3, $\sim$23.8 and $\sim$23.4~mag in the $grz$ bands respectively. The DECam based surveys typically show a double peaked imaging depth distribution \footnote{See https://www.legacysurvey.org/dr10/description/}. The deeper peak is associated with the DES survey with  5$\sigma$ point source depths of $\sim$25.3, $\sim$25.0 and $\sim$23.9~mag, and the shallower peak is mostly associated with the DECaLS survey with depths of $\sim$24.8, $\sim$24.2 and $\sim$23.3~mag ($grz$).

The ACT-DR5 footprint ranges in Declination from -60 to +20 and therefore lies nominally in the southern part of the LS. While recognizing the fact that additional Legacy Survey North data are present within the footprint, we do not use those data, because they do not expand the coverage of the ACT-DR5 footprint. 

\subsection{WISE}\label{sec:unwise}
To expand the redshift reach of our cluster search we follow our previous work on SPT tSZE-selected clusters \citep{Klein24a,Bleem24} and use WISE to identify high redshift clusters. 
The WISE satellite is a mid-infrared telescope with a main mirror of 40\,cm observing in four bands at 3.4\,\textmu m, 4.6\,\textmu m, 12\,\textmu m and 22\,\textmu m (referred to as $w1$, $w2$, $w3$, $w4$, respectively).
The WISE observation campaign can be divided into three phases. During the main phase there was sufficient coolant to observe the entire sky 1.5 times in all four bands. Immediately following the main campaign was a second phase known as NEOWISE, which completed the second full-sky observations in the w1 and w2 bands, where coolant was not required. A third phase of WISE observations (NEOWISE-R) began in September 2013 after WISE was reactivated following more than two years of inactivity. Since then, WISE has been conducting a full-sky survey approximately every six months.

In our study, we utilize the most recent version of the unWISE catalog \citep{unWISE19} that is based on the nine-year full-depth unWISE coadds \citep{unWISE9y}. This version of the catalog incorporates all WISE data up until December 13 2021. It was constructed using the unblurred coadds of WISE imaging data \citep[unWISE][]{unWISE1,unWISE9y} and incorporates enhanced source detection and deblending modeling for densely populated areas.

\section{Confirming ACT-DR5 clusters}

To confirm clusters of galaxies in the ACT-DR5 based candidate list, we follow a similar approach to our previous work on SPT tSZE-selected clusters \citep{Klein24a,Bleem24} by running two instances of the MCMF cluster confirmation tool \citep[see details in][]{Klein18,Klein19}. The first instance uses optically selected galaxies from LS-DR10 and is the version used in our work on RASS X-ray-selected clusters \citep{Klein23}. The availability of WISE photometry in LS-DR10-- as with our work on SPT tSZE-selected clusters using DES-- allows us to expand the redshift reach of this instance of MCMF out to z=1.5. 
The second instance of MCMF, which we call the high-z extension, is based on IR selected galaxies from the unWISE catalog. Compared to LS-DR10 the IR selection allows one to identify massive galaxies to higher redshift, which enhances the capability to confirm clusters at the high redshift end of the sample. Given our experience with SPT tSZE-selected clusters we expect that the vast majority of ACT-DR5 clusters will be confirmed with our main instance of MCMF using LS-DR10, while only $\sim5-10$\% of the sample will be confirmed by our high-z extension.
Because in this work we closely follow our previous work on confirmation of ICM selected clusters, we will only give a brief overview of MCMF in the following subsections and refer the interested reader to the additional details provided in our previous work \citep{Klein23, Klein24a}.

\subsection{MCMF}\label{sec:mcmf}
The MCMF algorithm was created to validate and describe cluster candidates identified in large X-ray or tSZE surveys. MCMF has proven successful in analyzing SPT tSZE-selected sources within the DES footprint \citep{Klein24a,Bleem15}. Furthermore, it has been utilized in conjunction with the LS-DR10 dataset to establish the optically-confirmed X-ray cluster catalog covering the entire sky \citep[RASS-MCMF;][]{Klein23}. Additionally, it played a crucial role in the optical follow-up of the initial eROSITA-based galaxy cluster catalog in the eFEDS early mission test field \citep{Klein22}. Moreover, it has been employed in the examination of new S/N>3 Planck tSZE-selected catalogs across the DES region \citep[MADPSZ][]{Hernandez22}. Throughout these various applications, enhancements and expansions have been made to the original method. The newly generated MCMF-based cluster catalogs have significantly increased the number of clusters that were previously identified in the same X-ray or tSZE datasets and subsequently analyzed through individual imaging and spectroscopy.
In addition to extending the samples, the MCMF method allows one to control the level of contamination in the new cluster samples.

The MCMF algorithm incorporates a red sequence technique \citep{gladders00,Rykoff14} that utilizes redshift and magnitude dependent color filters, such as $g-r$, $r-z$, $z-w1$.
It also employs radial weighting using a projected NFW profile centered at the tSZE or X-ray selected candidate position, as well as a specific magnitude range depending on redshift. It also incorporates a tSZE- or X-ray-based mass proxy to estimate $R_{500}$ for a given redshift z.
For each cluster candidate, the color and radially weighted, background-subtracted richness $\lambda(z)$ within R500 is calculated as a function of the candidate redshift, which is initially unknown. The peaks in $\lambda(z)$ are then identified and modeled with so-called ``peak profiles'', which is described below. 
Examples of peak profiles and their best fit to $\lambda(z)$ profiles of clusters are presented in previous MCMF analyses \citep[e.g. Figs. 4 \& A2,][]{Klein19}. These peaks with associated richnesses and redshifts are then collected and processed further as described in the following subsections.
Note that the peak profile models are built using renormalised stacks of individual $\lambda(z)$ profiles from real clusters with spectroscopic redshift measurements (spec-z's). This approach ensures that no additional correction between MCMF and spec-z redshifts are needed.

To confirm cluster candidates, we assess the probability that a given optical counterpart is a coincidental overlap rather than a genuine counterpart to the ICM-selected cluster candidate. This assessment requires knowledge of the typical distribution of contaminants based on their richness and redshift within the survey region. To obtain this information we supplement the analysis that we do on the candidates positions with an analysis of random positions within the tSZE or X-ray survey footprint. In the ACT-DR5 analysis, these random positions have a similar density and SNR distribution as the candidate clusters but exclude regions with real ACT candidates. The richness distribution of these random positions provides an estimate of the non-tSZE detected structures (noise, projections, undetected clusters) within the candidate list.  

To be able to control the contamination of the final ACT-DR5~MCMF cluster sample, we calculate a quantity \fcont, where ''cont'' is short for contamination.  High values of \fcont\ indicate a higher probability that the candidate in question is a chance superposition rather than a real cluster.  \fcont\ is calculated using the richness distributions along the random lines-of-sight $f_\mathrm{rand}(\lambda,z)$ and the richness distributions $f_\mathrm{obs}(\lambda,z)$ towards the candidates at each  redshift $z$. 

That is, for each candidate $i$ we calculate the number of random lines-of-sight within a redshift bin with richness $\lambda\ge\lambda_i$ and divide by the number of tSZE candidates within the same redshift bin with  $\lambda\ge\lambda_i$. This ratio is then re-scaled according to the total number of SPT candidates and random lines-of-sight.
\begin{equation}\label{eq:fcont}
   f_{\mathrm{cont}}(\lambda_i,z_i)=\frac{\int_{\lambda_i}^{\infty} f_\mathrm{rand}(\lambda,z_i) d\lambda}{\int_{\lambda_i}^{\infty}
f_\mathrm{obs}(\lambda,z_i) d\lambda},
\end{equation}
This \fcont\ parameter is calculated for each richness peak associated with a candidate.  The peak showing the lowest value of \fcont\ is assigned as the best optical counterpart for the SPT-SZ candidate, because it is the most likely to be a real cluster. 

The cluster sample itself can then be defined as those candidates showing an \fcont\ below a certain threshold value $f_\mathrm{cont}^\mathrm{max}$. The threshold value corresponds to the fraction of the contamination in the initial candidate list that makes it into the final cluster catalog at each redshift $z$.
The contamination of the resulting final cluster catalog would then be $f_\mathrm{tSZE-cont} \times f_\mathrm{cont}^\mathrm{max}$, where the confirmed catalog contains all candidates with $f_{\mathrm{cont}}(\lambda_i,z_i)\le f_\mathrm{cont}^\mathrm{max}$ and the initial contamination of the candidate catalog is $f_\mathrm{tSZE-cont}$.
As an example, if the input catalog is known to be 50\% pure and an \fcont\ threshold value $f_\mathrm{cont}^\mathrm{max}=0.2$ is employed, then the contamination fraction of the confirmed cluster catalog would be $0.5\times 0.2=0.1$ or 10\%.

\subsubsection{MCMF on Legacy Survey DR10}\label{sec:MCMFonLSDR10}
The version of MCMF applied here is-- aside from the different mass proxy-- largely the same as the version applied to X-ray cluster candidates from the 2RXS catalog \citep{Boller16} to create the RASS-MCMF catalog \citep{Klein23}. 
The most significant change between the MCMF version used to create RASS-MCMF and this work is that we exclude the $w1-w2$ color in our baseline approach, and we recalibrate the scatter around the red sequence model in the $z-w1$ color. This results in a better behaved redshift evolution of the richness-mass relation.  These changes imply that the there are systematic differences between our ACT-DR5~MCMF and RASS-MCMF richnesses for clusters of similar redshift and mass. We therefore provide a table to transform RASS-MCMF richnesses to our new baseline richnesses in the appendix. We provide a more detailed comparison between RASS-MCMF and our new ACT-DR5 catalog below in Section~\ref{sec:ACTxRASS}.

In addition, following the approach we used in our work on SPT tSZE selected candidates \citep{Klein24a,Bleem24}, we improve the estimate of the richness distributions along random lines-of-sight with a second iteration. The richness distribution along random lines-of-sight $f_\mathrm{rand}(\lambda,z)$ is supposed to resemble the expected richness distribution of contaminants as a function of redshift. Given the correlation between the tSZE observable SNR (often referred to as $\hat\zeta$ or $\xi$ in our previous SPT analyses), cluster mass, and likelihood of a source being a real cluster, the initial choice of using the identical SNR distribution as the full candidate list causes the estimate of \fcont\ to be mildly biased high. To avoid this bias we use the SNR distribution of rejected candidate systems (\fcont>0.3) as a proxy of the distribution of contaminants.  We select a subsample of random lines of sight that follows this tSZE observable distribution and use that to remeasure \fcont\ for all candidates.

\subsubsection{MCMF on unWISE}\label{sec:HIGHZ}
In addition to passive galaxies becoming fainter as redshift and distance increase, the spectral range covered by the DECam bands no longer encompasses the 4000,\AA \ break at redshifts $z\gtrsim1$. Consequently, photometric redshift estimation becomes increasingly uncertain at these redshifts.
For confirming high-redshift clusters and improving photometric redshift estimates, transitioning to redder bands like the near- or mid-IR range covered by the Spitzer or WISE observatories proves advantageous. Previous studies have utilized data from both observatories for high-redshift cluster searches \citep{Muzzin09,Madcows} and cluster validation \citep{Bleem15,Bleem20,Bleem24,Klein24a}. In this work we follow closely the approach presented in \citet{Klein24a}.

In our current analysis, we primarily rely on a version of the unWISE catalog \citep{unWISE3cat}, which incorporates eight years of $w1$ and $w2$ band WISE imaging data from the NEOWISE-R phase. WISE data cover the entire sky, enabling the matching of unWISE and LS-DR10 catalogs for optical+IR photometry of all WISE sources. Optical to WISE galaxy colors (e.g., $z$-$w1$) exhibit strong redshift dependence, making them suitable for obtaining high-quality cluster photometric redshift estimates. 

However, a limitation of WISE $w1$ is its large PSF ($\sim6"$), which becomes problematic in regions with high source density like the cores of galaxy clusters. In such dense areas, distinguishing individual sources and deblending fluxes pose challenges. Here, the improved modeling of crowded regions in unWISE compared to the previous ALLWISE catalog becomes important.

Finally, we accommodate masks or missing data in different surveys by deriving separate richness estimates for cluster regions with different coverage (LS-DR10 + $w1$ only, WISE $w1$ only, WISE $w1$, $w2$ \& LS-DR10) and summing them for the final cluster richness estimate. With this approach, we need only track the masked area in $w1$ imaging. The total richness in the high-redshift code is therefore given as
\begin{align}
\label{eq:richnto}
\lambda_{\mathrm{HZ}}(z)=&\lambda_{\mathrm{LSDR10+w1}}(z)+\lambda_{\mathrm{LSDR10+w1+w2}}(z)+\lambda_{\mathrm{w1+w2}}(z)+\lambda_{\mathrm{w1}}(z),\nonumber
\end{align}
where the individual richnesses are defined in the same manner as the standard MCMF richness \citep[see][]{Klein18,Klein19}, with the color-weights depending on the availability of the bands ($i,z,w1,w2$ for $\lambda_{\mathrm{LSDR10+w1+w2}}$, $i,z,w1$ for $\lambda_{\mathrm{LSDR10+w1}}$, $w1$, $w2$ for $\lambda_{\mathrm{w1+w2}}$ and no color weight for $\lambda_{\mathrm{w1}}$).

The high-$z$ cluster confirmation code has been applied to all candidates over the redshift range $0.63<z<2.0$. Similar to the optically-based MCMF code, runs are performed along random lines-of-sight, and \fcont\ is calculated for potential counterparts to each tSZE candidate. Clusters with spectroscopic redshifts available for a subset of the sample are utilized to calibrate the WISE-based measurements. Moreover, the overlap in redshift between the optically-based MCMF and the high-z WISE-based run of $0.63<z<1.5$ enables the comparison of richnesses.

\subsubsection{Spectroscopic redshifts}
\label{sec:spectroscopicredshifts}
We follow the same approach as adopted in \cite{Klein23} and estimate spectroscopic redshifts for the best optical counterpart identified using MCMF by employing public spectroscopic galaxy redshifts. For this analysis we use the same collection of spectroscopic redshifts as was used in \cite{Klein23} that mainly includes redshifts from the SDSS~DR17 \citep{sdss17}, 2dFGRS \citep{2df}, 6dFGS \citep{jones09}, 2MRS \citep{2MRS} and GLADE+ \citep{GLADE}. Additionally, we match to known clusters with spectroscopic redshifts.
We first match the spectroscopic catalog with the BCG positions using a maximum positional offset of 2~arcsec. We then search for all spectroscopic galaxies within 2 Mpc around the ACT position and $\vert z_\mathrm{cluster}-z_\mathrm{spec} \vert < 0.025(1+z_\mathrm{cluster})$.
From this list of potential spectroscopic member galaxies, we derive the median redshift and finally derive the cluster redshift using all galaxies within $\left| \delta z\right|<0.015$ from the median redshift. If a BCG spectroscopic redshift exists, we instead select galaxies within $\left|\delta\right| z<0.015$ from the BCG redshift. In our final cluster catalog, we only list spectroscopic redshifts that have been derived from at least two members, that has a BCG redshift or that has matches to previously known clusters with spectroscopic redshifts. 

\begin{figure}
\begin{center}
\includegraphics[width=0.99\linewidth]{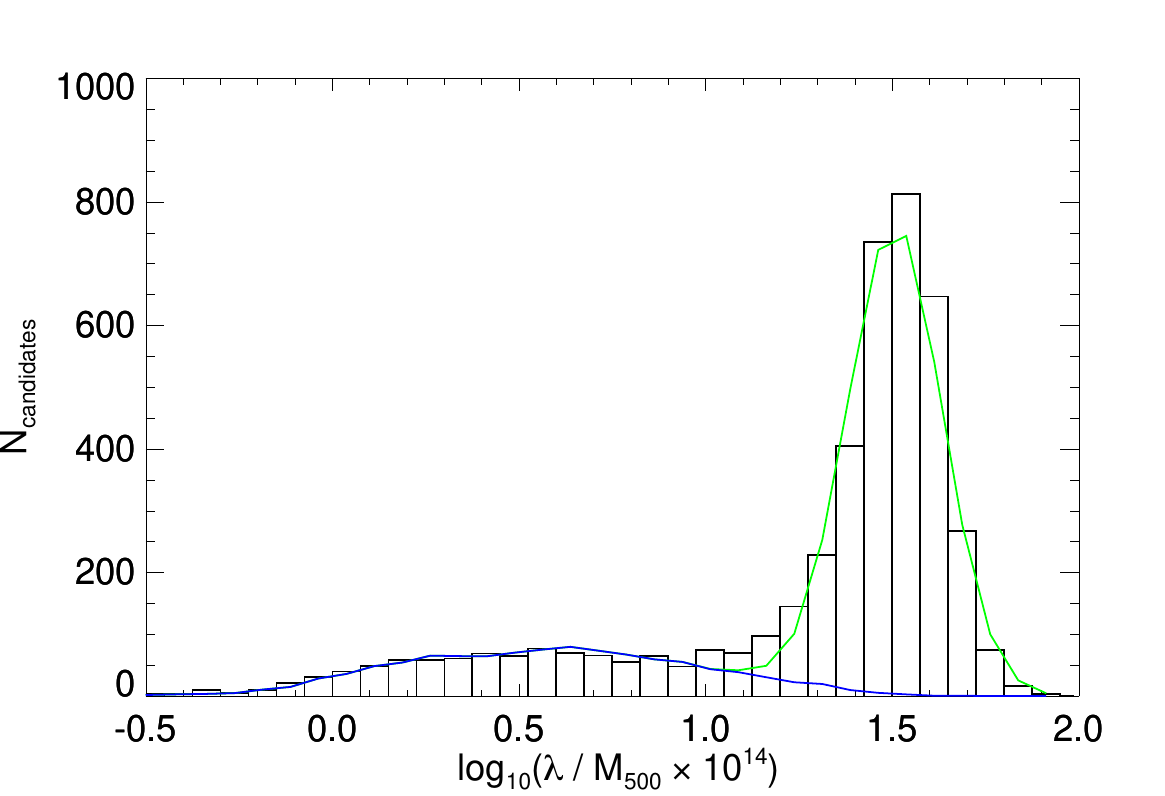}
\caption{Measured distribution of ACT-DR5 candidates with SNR>4.5 as a function of $\log_{10}(\lambda/M_{500})$. At low values the distribution is dominated by contamination caused by noise fluctuations in the mm-wave maps, which are well described by the contamination model (blue), while the high values are dominated by true galaxy clusters. The joint contamination plus cluster distribution is shown in green and blue.}
\label{fig:nfalsecandmethod}
\end{center}
\end{figure}

\subsection{Contamination in the ACT-DR5 candidate list}\label{sec:initialconta}
\begin{figure}
\begin{center}
\includegraphics[width=0.99\linewidth]{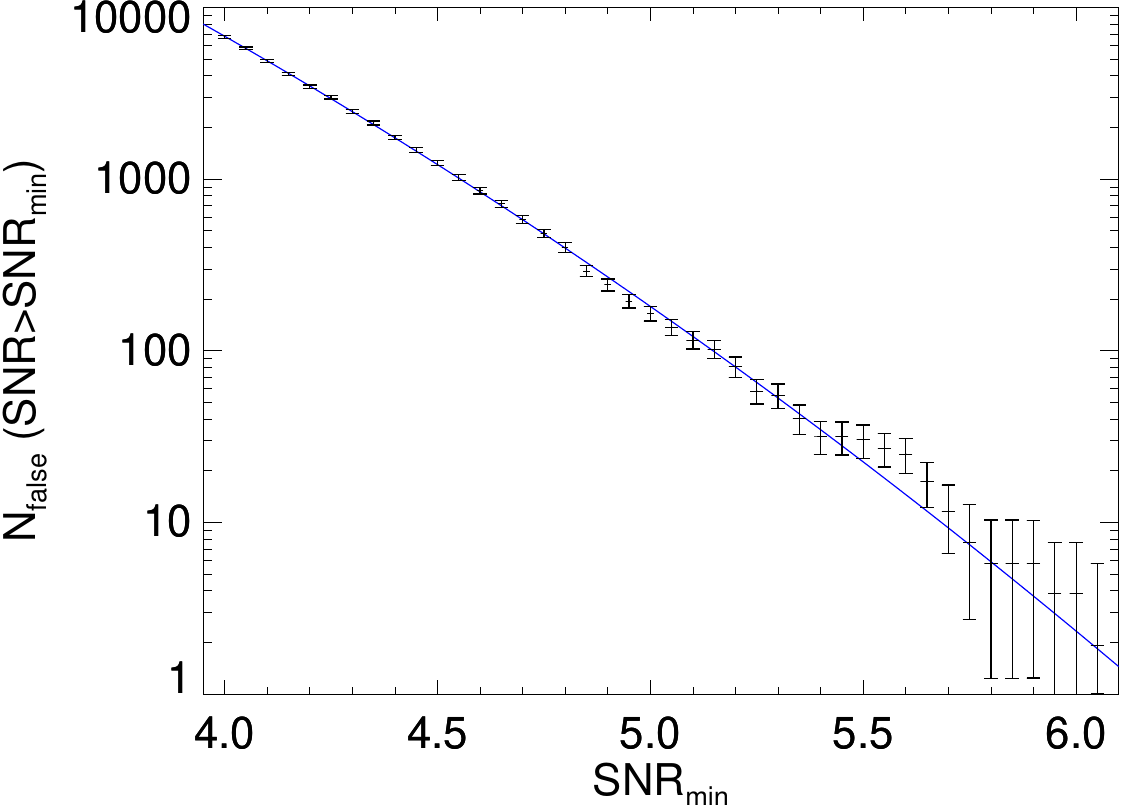}
\includegraphics[width=0.99\linewidth]{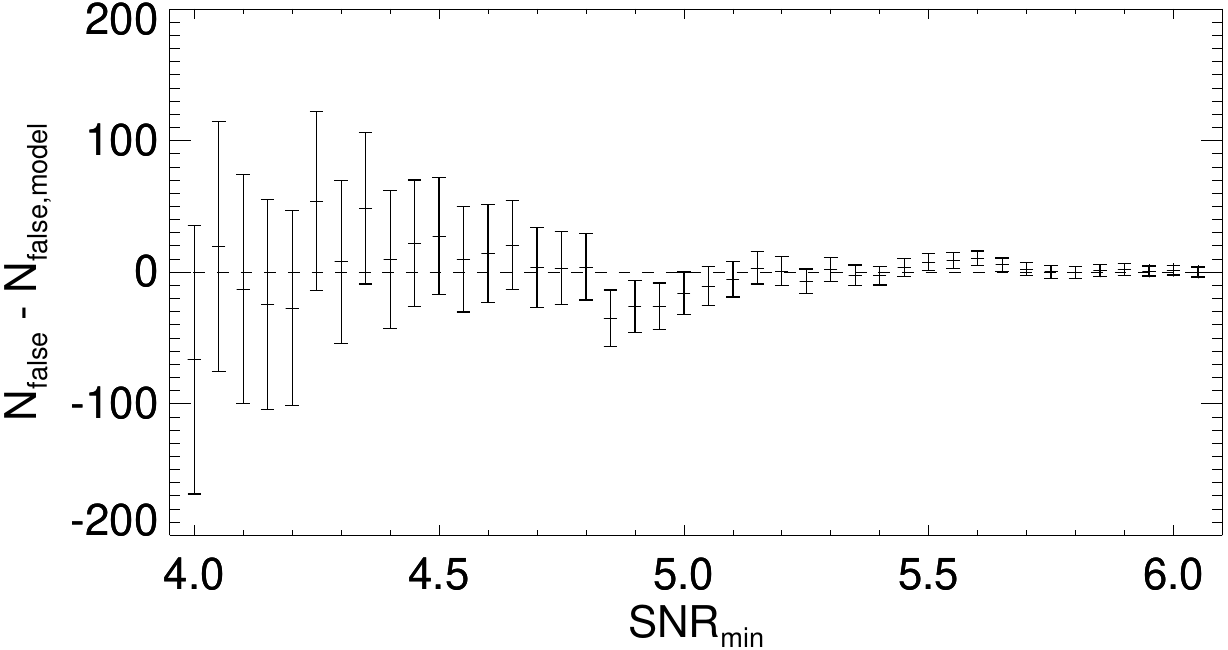}
\caption{Measured cumulative number of false detections from noise fluctuations in the mm-wave maps as a function of the tSZE SNR threshold. The blue line shows the best fit Gaussian noise model. The two parameter fit describes the data over full regime explored, corresponding to almost four orders of magnitude.}
\label{fig:nfalsecandidates}
\end{center}
\end{figure}

Systematic optical and IR follow-up of ACT-DR5 candidates with MCMF not only allows one to confirm clusters, differentiating between likely clusters and noise fluctuations, but it also allows one to statistically estimate the total number of clusters and noise fluctuations in the sample \citep{Klein22,Hernandez22,Klein23,Klein24a}. 
One does this by using the fact that the richness and redshift distributions along random lines of sight are representative of the distribution of noise fluctuations in the candidate list. The accuracy of the estimate improves when there is a region in the parameter space that is dominated by noise fluctuations. In cases where the ICM-based mass proxies are noisy or not reliable 
it is most robust to use solely the richness distribution of the sample selected along random lines of sight and to employ the sources with low richness to estimate the total number of contaminants \citep[][]{Hernandez22,Chiu23}. For cases with meaningful mass proxies, such as in this work or in our previous work using SPT-SZ or eROSITA eFEDS extent-selected clusters \citep{Klein22, Klein24a}, one can model the distribution in $\log_{10}(\lambda/M_{500})$ space, where $M_{500}$ is the ICM-based mass estimate derived from a calibrated observable-mass relation. 
Given that the mass trend of the richness-mass relation $\lambda_{M500}\sim1$, moving to $\log_{10}(\lambda/M_{500})$ space causes the cluster population to be projected onto one location (value) with some scatter around that value due to scatter in the richness-mass relation. In this space the clusters and contaminants tend to be better separated. In addition,  this space is a well suited for building a more sophisticated model of the full population that includes a contamination component calibrated using systems identified along random lines of sight, and a cluster model that we approximate using a log-normal distribution.

In this work we follow closely our analysis of SPT tSZE-selected cluster candidates and fit a population model described by four parameters: three parameters for the log-normal distribution describing the true clusters and one for the relative normalisation of contaminant distribution. In Fig~\ref{fig:nfalsecandmethod} we show an example for the ACT-DR5 candidates with SNR>4.5. Using this method, we measure the number of contaminants as a function of SNR threshold SNR$_\mathrm{min}$. Given that the cluster candidate search was performed in the 90 and 150\,Ghz bands where the tSZE signature appears as a negative signal, we expect the only true or physical sources to be clusters, while the contamination is solely caused by noise in the mm-wave maps. Given the noise characteristics, the number of noise fluctuations as a function of SNR$_\mathrm{min}$ should be well described by a Gaussian noise model with only two free parameters: the standard deviation $\sigma $and the amplitude $A$, with a mean set to 0. The number of false detections then is
\begin{equation}\label{eq:nfalse}
N_\mathrm{false}=A\int_{\mathrm{SNR}_\mathrm{min}}^{\infty}\frac{1}{\sqrt{2\pi\sigma^2}}e^{-\frac{x^2}{2\sigma^2}} dx.
\end{equation}
In this case the standard deviation roughly corresponds to the noise of the map used for source detection. Although the analysis is done in terms of SNR, we do expect some mild offset from the expected value from unity, even in the case of a correct estimate of the noise properties of the map. This bias is caused by the fact that the detection pipeline tries to maximize the SNR at a given position by adopting different filter sizes (see Sec.\ref{sec:ACTdata}). The amplitude of the Gaussian model is given by the effective number of independent detection cells. It therefore scales with survey area and the size of the matched filter used for the detection. As clusters are only barely resolved in ACT one can expect that the size of the filter (or detection cell) is close to the beam size of ACT, our expectation value for the amplitude therefore is close to the number of beams that fit into the ACT footprint.

In Fig.~\ref{fig:nfalsecandidates} we show the number of false detections as a function of SNR$_\mathrm{min}$ together with the best fit Gaussian noise model. As is clear, the model describes the data remarkably well over almost four orders of magnitude in dynamical range. 
The best fit amplitude of the Gaussian model is $A=(28.1\pm1.4) \times 10^6$. Using the size of the footprint this cell area corresponds to 1.7\,arcmin$^2$, which is, as expected, close but mildly larger than the area covered by the beam at 150\,GHz of $\sim1.5$\,arcmin$^2$.
The measured standard deviation $\sigma=1.146\pm0.004$ is, as expected, close but significantly larger than the naive assumption of $\sigma=1$. 

Using our best fit model and the number of candidates, we estimate the fraction of contaminants in the candidate list to be $53.7\pm0.5$\%. We note that this result is robust against the particular choice of method to model the population of noise fluctuations in the candidate list. Repeating the same exercise using the method applied in the MCMF analysis of the Planck cluster candidate list \citep{Hernandez22}, where the method uses richness and constrains the contamination in the contamination dominated regime ($\log_{10}(\lambda/M_{500})<0.7$), yields contamination fractions within 0.5\% of those we estimate with our baseline analysis. 

\begin{figure}
\begin{center}
\includegraphics[width=0.99\linewidth]{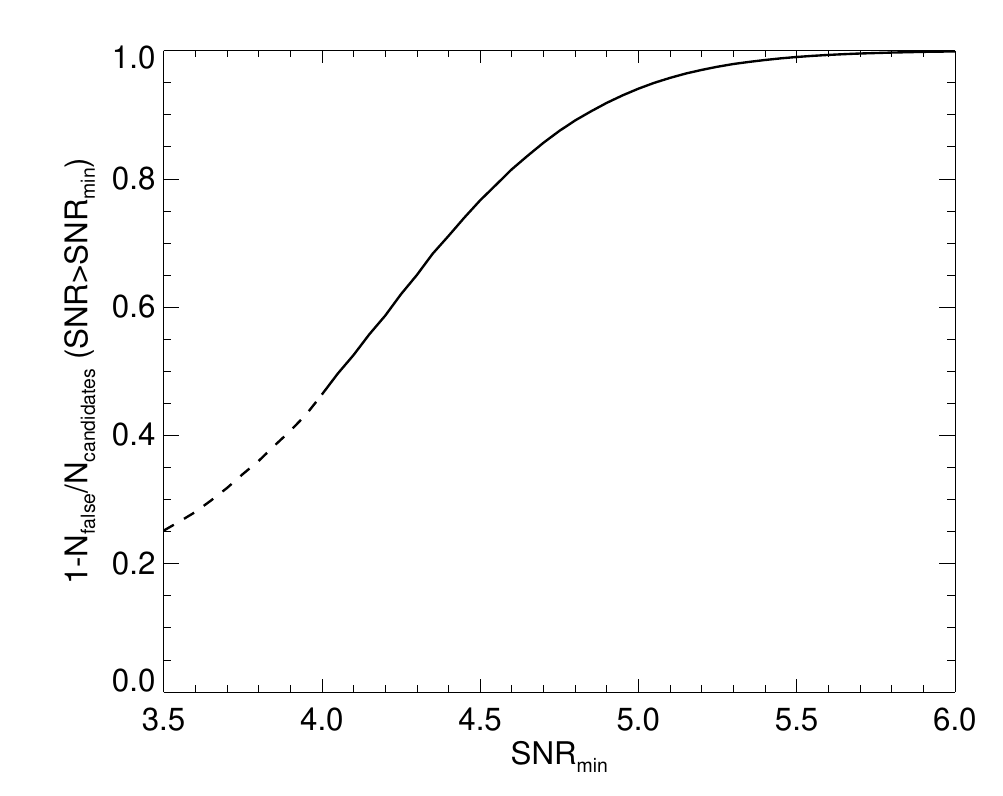}
\caption{Measured purity of the tSZE selected candidate list as function of SNR threshold.  The ACT-DR5 candidate list selected with SNR>4.0 has a purity of 53.5($\pm$0.5)\%.}
\label{fig:candidatelistpurity}
\end{center}
\end{figure}

With the model describing the number of contaminants and the number of observed candidates we can measure the purity ($1-N_\mathrm{false}/N_\mathrm{candidates}$) as a function of SNR threshold SNR$_\mathrm{min}$, as shown in Fig.\ref{fig:candidatelistpurity}. Assuming that the Gaussian noise model holds also at lower SNR$_\mathrm{min}$<4.0, we show the expected purity of the candidate list down to lower SNR$_\mathrm{min}$ (dashed line), which yields a purity of only 25\% at SNR$_\mathrm{min}=3.5$.
In the SNR range between 3.5 and 4.0 we would expect $\sim4,800$ additional clusters, but at the same time adding $\sim25,000$ noise fluctuations. Lowering the threshold therefore significantly increases the amount of contamination, which would require stricter optical cleaning with MCMF.  The impact of optical cleaning on the over-all cluster sample would become more important, and uncertainties in that cleaning would begin to play an important role in the selection function.

\section{The ACT-DR5~MCMF cluster catalog}
In this section we define the cluster sample considering completeness and purity and report the basic properties of the sample.
With the estimate of the purity of the initial list of ACT-DR5 candidates, the total number of true clusters is known to be $5872\pm99$, 97\% of which have good optical and NIR data for confirmation. 
The MCMF-based \fcont\ estimator together with the initial level of contamination provides an estimate of the contamination fraction of the final \fcont-cleaned sample. When this information from the tSZE selection and optical cleaning is combined, it allows one to trace the completeness and purity of the \fcont-cleaned sample and provides guidance for the optimal sample selection.

\subsection{Definition of the ACT-DR5~MCMF cluster catalog}
\begin{figure*}
\begin{center}
\includegraphics[width=0.4\linewidth]{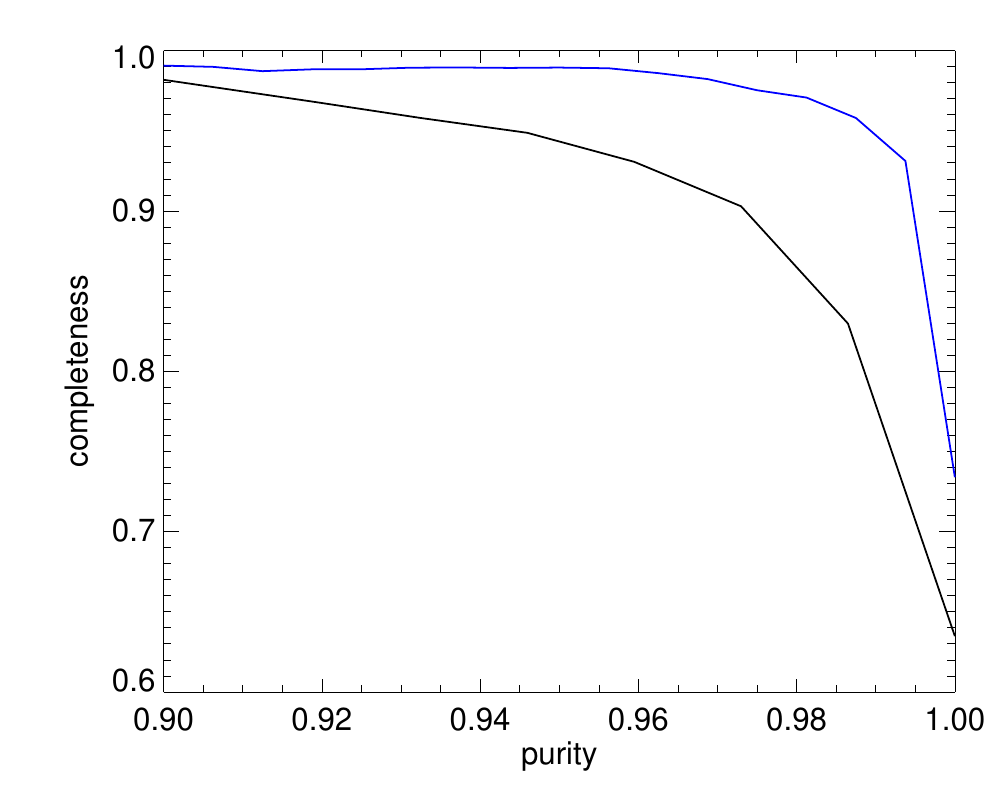}
\includegraphics[width=0.47\linewidth]{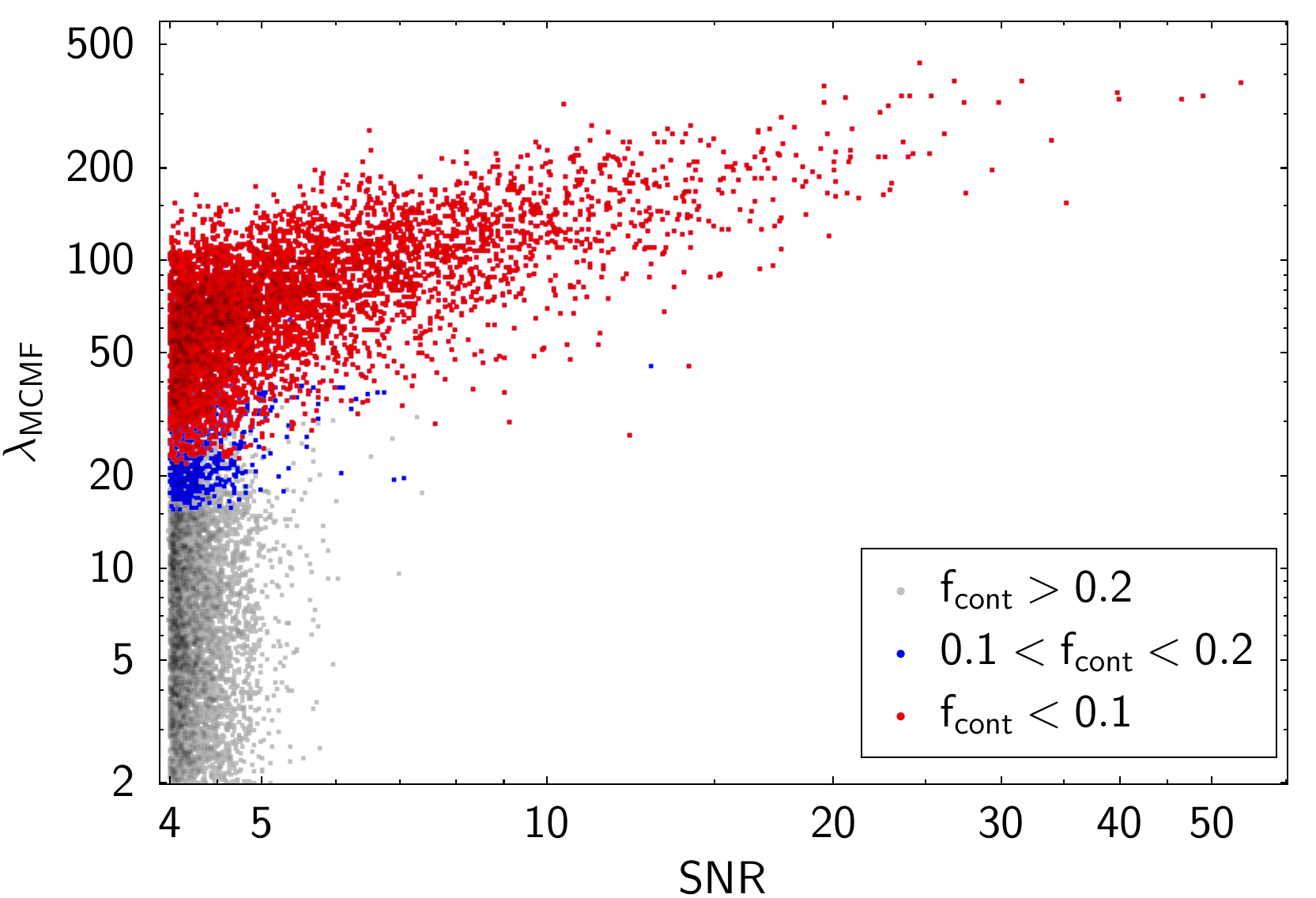}
\caption{Left: Completeness with respect to the initial tSZE selection versus purity after optical cleaning with \fcont\ ($f_\mathrm{cont}^\mathrm{max}$ increasing from left to right) for candidate lists imposing SNR>4 (black) and SNR>4.5 (blue). Right: Distribution of ACT-DR5 candidates in richness vs. SNR, color coded according to the \fcont\ ranges.}
\label{fig:complpurity}
\end{center}
\end{figure*}

We follow the approach used in the creation of the SPT-SZ MCMF selected cluster sample \citep{Klein24a} and explore the contamination fraction and completeness as a function of the $f_\mathrm{cont}^\mathrm{max}$ selection threshold.  If a cluster sample is created by imposing an upper limit in $f_\mathrm{cont}^\mathrm{max}$, then the corresponding contamination fraction of the sample is the product of $f_\mathrm{cont}^\mathrm{max}$ and the contamination fraction of the initial candidate list (i.e., $53.7\pm0.5$\% for the SNR>4 ACT-DR6 sample). In the left plot of Fig.~\ref{fig:complpurity} we show the completeness with respect to the initial tSZE selection and the purity achieved after imposing a $f_\mathrm{cont}^\mathrm{max}$ selection that decreases from left to right along each of the lines. On the right side of Fig.~\ref{fig:complpurity} we show the distribution of candidates in richness versus ACT-DR5 SNR. There is a clear correlation between richness $\lambda_\mathrm{MCMF}$ and tSZE SNR, even in the presence of variations in ACT-DR5 survey depth over the sky. The number of low-richness systems, which are likely noise fluctuations, significantly rises towards lower SNR, while the typical richness of likely true clusters decreases towards lower SNR. The boundary between the noise fluctuation dominated region and the real cluster dominated region is around $\lambda_\mathrm{MCMF}=20$. 

One rough rule of thumb for defining a cluster sample where the residual contamination is unimportant for most analyses is to ensure that the residual contamination is smaller than the expected Poisson noise, which for a sample of $\sim5,900$ clusters is about 1.3\%. For some analyses it is important to avoid a situation where a significant number of real clusters are lost due to the optical cleaning.
In terms of completeness this threshold is reached by imposing a cut of $f_\mathrm{cont}^\mathrm{max}=0.2$, which  corresponds to a purity of $\sim90$\% (i.e., contamination fraction is 0.2$times$0.53 in the case of the SNR>4 sample). This selection yields a sample with significant contamination but the impact of the optical confirmation is negligible in terms of removing real clusters. To obtain samples with negligible levels of contamination one either has to accept $\sim16$\% incompleteness introduced by the more aggressive cleaning implied by a dropping threshold in $f_\mathrm{cont}^\mathrm{max}$, or one has to create the cluster sample from a candidate list with a higher SNR threshold to increase the initial purity of the tSZE selected candidate list. As an example, imposing SNR$_\mathrm{min}=4.5$ increases the initial purity to 76\% allowing for a cluster sample with 98\% purity and completeness, which is close to the level of the Poisson noise for a cluster sample of this size.

Because all cleaner subsamples can be constructed from the $f_\mathrm{cont}^\mathrm{max}=0.2$ and SNR$_\mathrm{min}>4.0$ selected sample, we use these two selection thresholds in defining our ACT-DR5~MCMF cluster catalog. In Table~\ref{tab:sample} we show sample selection parameters and key numbers for the ACT-DR5~MCMF sample along with two example subsamples: the aforementioned sample with SNR$_\mathrm{min}=4.5$ and 98\% purity and completeness (SNR-4.5) and the Purity-95 sample that has increased purity and decreased completeness with respect to the full ACT-DR5~MCMF cluster sample. Following A\&A policies the catalog will be made available on CDS after acceptance of the paper. Earlier access to the catalog can be granted upon reasonable request. The description of the catalog entries can be found in Appendix~\ref{app:ACTMCMFcols}.

\begin{table}\caption{Properties of the ACT-DR5~MCMF cluster catalog along with two subsamples. The table shows sample name, the optical selection criterion $f_\mathrm{cont}^\mathrm{max}$, the tSZE selection criterion SNR$_\mathrm{ min}$, the expected final sample purity, the completeness with respect to the tSZE selection and the number of selected clusters $N_\mathrm{cl}$.}
    \label{tab:sample}
    \centering
    \resizebox{\linewidth}{!}{%
    \begin{tabular}{lccccc}
        \hline
        \hline
        \T &  &  &  Purity & Comp. \\
        \B Cluster Sample& $f_{\mathrm{cont}}^{\mathrm{max}}$ & SNR$_\mathrm{min}$ & $[\%]$ & $[\%]$ & $N_\mathrm{cl}$ \\
        \hline
        \T ACT-DR5~MCMF & 0.2 &  4.0  &  89.7 & 97.8 & 6237\strut\\
        Purity-95 & 0.1 & 4.0 & 94.6 & 93.1 & 5604  \\
        \B SNR-4.5 & 0.1 & 4.5 & 97.6 & 98.2 & 3747 \\
         \hline
    \end{tabular}
    }
\end{table}


\subsection{Properties of the ACT-DR5~MCMF cluster catalog}
\begin{figure}
\begin{center}
\includegraphics[width=0.99\linewidth]{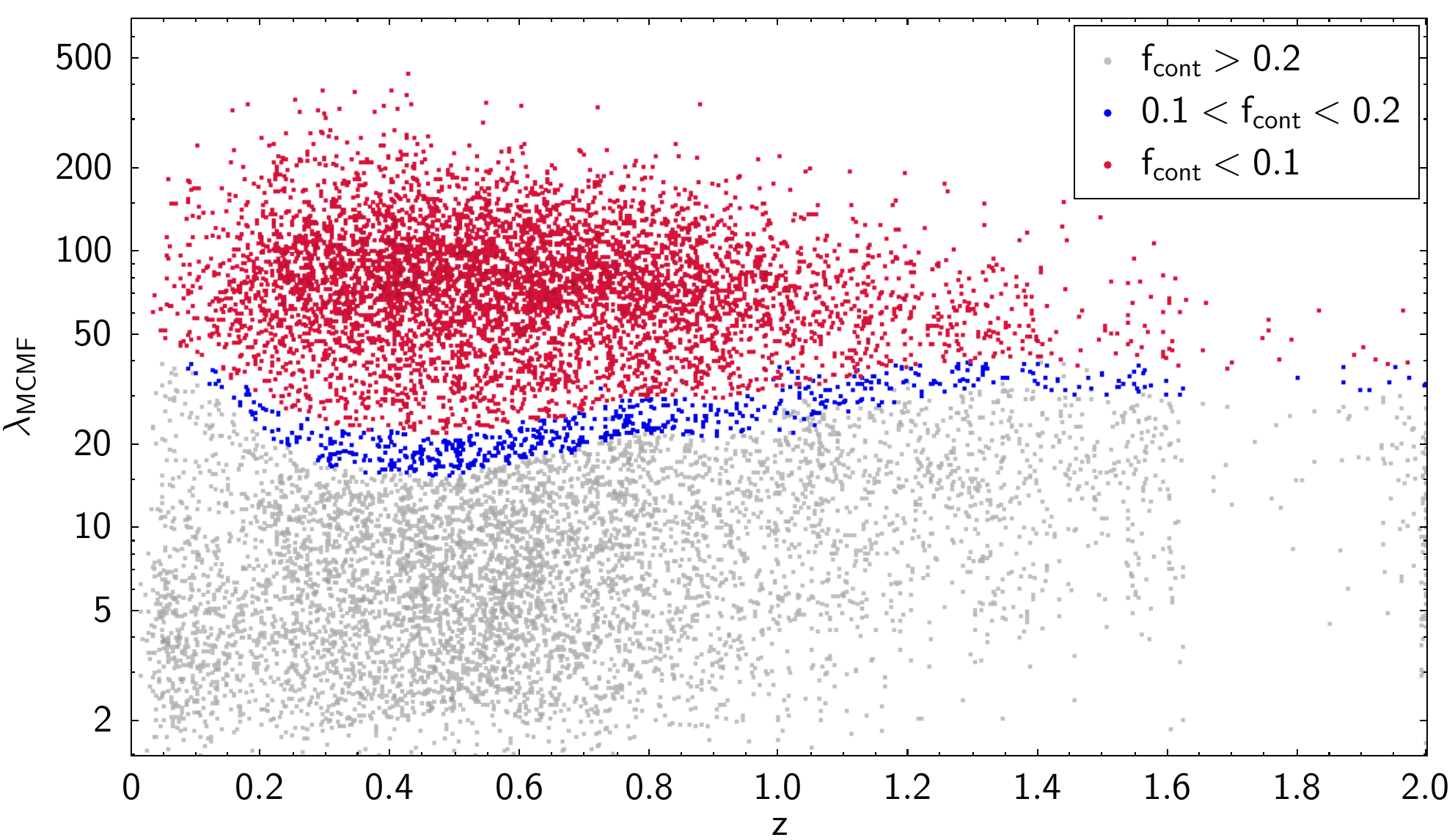}
\caption{ACT-DR5 candidates distributed in richness vs. redshift space. Color coded are unconfirmed candidates in grey, confirmed systems with 0.1<\fcont<0.2 in blue and confirmed clusters with \fcont<0.1 in red.}
\label{fig:ACTrichvsred}
\end{center}
\end{figure}
\begin{figure*}
\begin{center}
\includegraphics[width=0.49\linewidth]{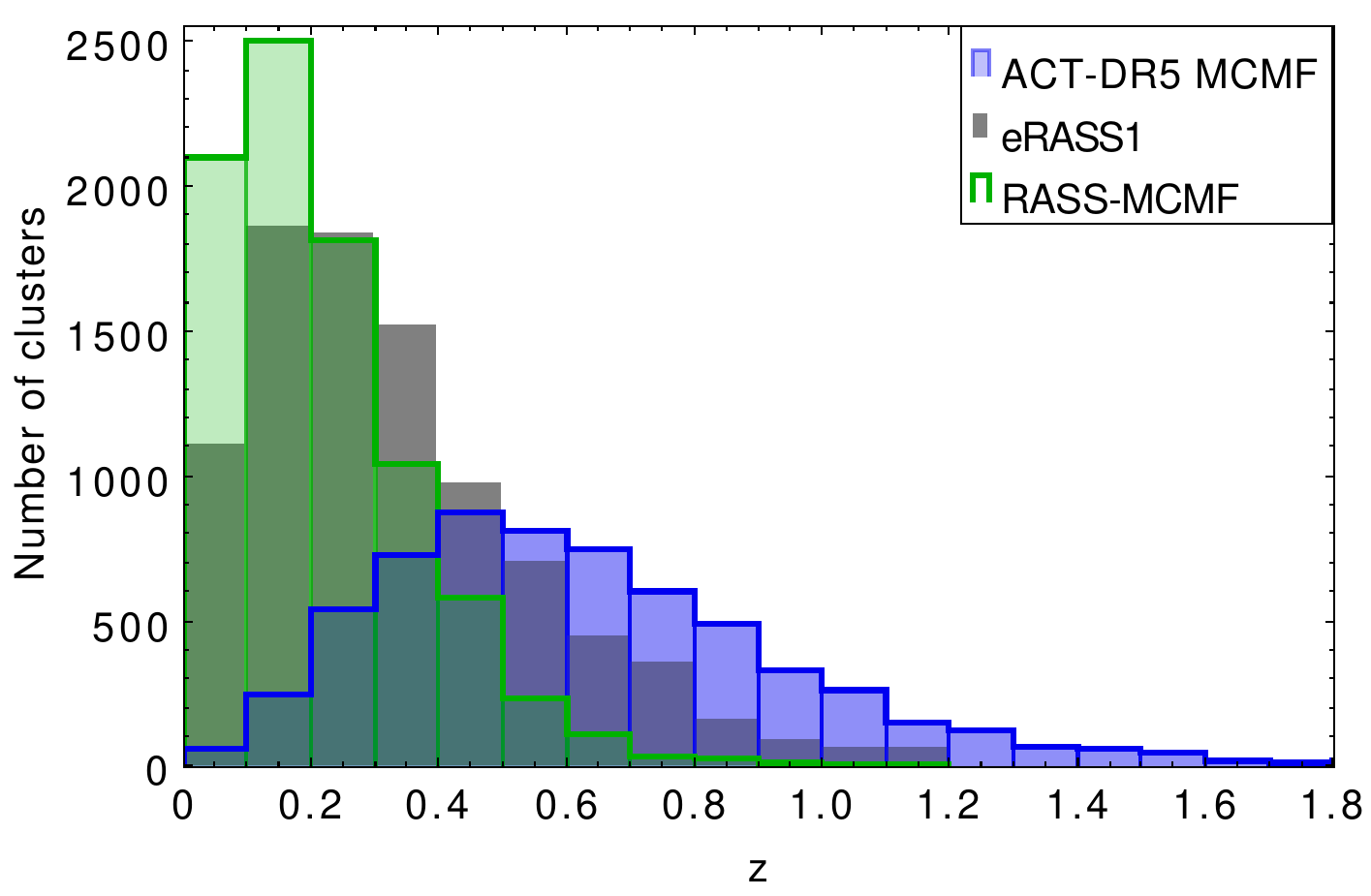}
\includegraphics[width=0.475\linewidth]{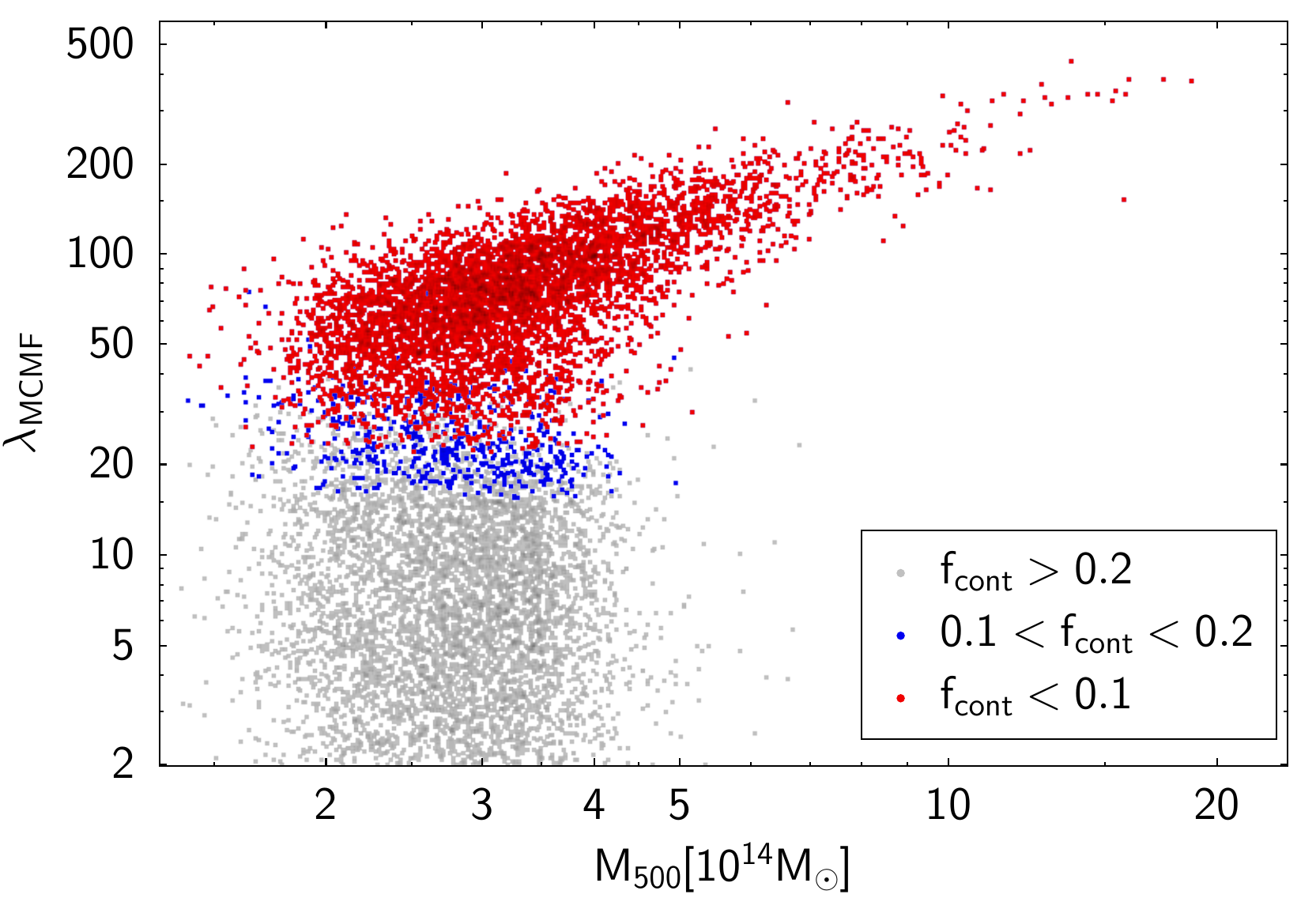}
\caption{Left: Redshift distribution of the three largest ICM-selected cluster catalogs to date. While RASS-MCMF provides the largest number of clusters at low redshift ($z<0.3$), ACT-DR5~MCMF provides the most clusters at high redshift ($z>0.5$). For the intermediate redshift range ($0.3z<0.5$) eRASS1 provides the largest number of clusters. Right: Richness versus tSZE-based mass of ACT-DR5 candidates and the ACT-DR5~MCMF sample. Color coded are unconfirmed candidates in grey, confirmed clusters with 0.1<\fcont<0.2 in blue and confirmed clusters with \fcont<0.1 in red.}
\label{fig:ACTredshift}
\end{center}
\end{figure*}

The properties of the ACT-DR5~MCMF cluster catalog and the two example subsamples are governed by two selections: the tSZE selection threshold in SNR and the \fcont\ selection that results in a redshift dependent richness cut. The SNR selection not only impacts the purity of the tSZE-based candidate list, but also- due to its correlation with mass- impacts the mass range of the selected clusters. Because richness also scales with cluster mass, the completeness with respect to the tSZE selection therefore increases at fixed richness threshold for increasing SNR thresholds. The SNR-45 sample therefore benefits two-fold from the increased SNR threshold, in purity and in completeness. A downside is that the SNR-45 sample is missing a significant number of low mass clusters compared to the default ACT-DR5~MCMF cluster sample. Due to the growth of structures over time this increase in effective mass selection is expected to impact the number of high redshift clusters more than low redshift clusters.

The MCMF method is designed to maintain the fractional contamination of the cluster sample constant at all redshifts. The redshift dependent richness cut implied by the \fcont\ selection depends to two things: the original tSZE selection and the properties of noise fluctuations and clusters in richness-redshift space. In Fig.~\ref{fig:ACTrichvsred} we show the distribution of ACT-DR5 SNR>4 candidates in richness-redshift space, highlighting two \fcont-selections in blue and red. The shape in the selection seen in this figure can be explained as follows. The upturn at low redshift is due to the properties of the tSZE selection (see Fig.7 in \cite{ACTDR5}) which leads to few candidates in this region. The lowest value is reached at $z\approx0.45$ where the number of real clusters reaches its maximum.  Beyond that redshift the richness threshold increases due to the limitations of the optical data to disentangle clusters and contaminants. For reasonably small values of \fcont\ ($<0.3$) the shape of the redshift dependent richness selection remains approximately unaltered and the difference in selection can be approximated by an overall shift in richness threshold.

Due to the redshift insensitivity of the tSZE based cluster detection, the ACT-DR5~MCMF cluster catalog contains a large number of clusters at high redshift. The median redshift of the sample is $z=0.58$.  In the left panel of Fig.~\ref{fig:ACTredshift} we show the redshift distributions of the three largest ICM-selected cluster samples to date. While the all-sky X-ray based RASS-MCMF cluster catalog \citep{Klein23} yields the largest number of clusters at $z<0.2$, the roughly half-sky ACT-DR5~MCMF cluster catalog provides the overall largest number of clusters at $z>0.6$. The largest current ICM-selected cluster catalog is the X-ray based eRASS1 \citep{Bulbul24} catalog. While covering a similar sky area as ACT-DR5, it provides the largest number of clusters in the intermediate redshift regime $0.2<z<0.5$.

In the right panel of Fig.~\ref{fig:ACTredshift} we show the distribution of richness versus a tSZE-based mass estimate for the ACT-DR5 candidates as well as for the official ACT-DR5~MCMF and the Purity-95 samples. Compared to the raw tSZE SNR shown in Fig.~\ref{fig:complpurity}, we see a tighter richness-mass relation for ACT-DR5~MCMF clusters. This property of a tSZE selected sample contrasts with an X-ray selected sample, where the X-ray count rate mass proxy scatters much more about the mean observable--mass relation \citep[][Singh et al., in prep.]{eRASS1cosmo}.

\subsection{High-z ACT-DR5~MCMF clusters}
\begin{figure*}
\begin{center}
\includegraphics[width=0.23\linewidth]{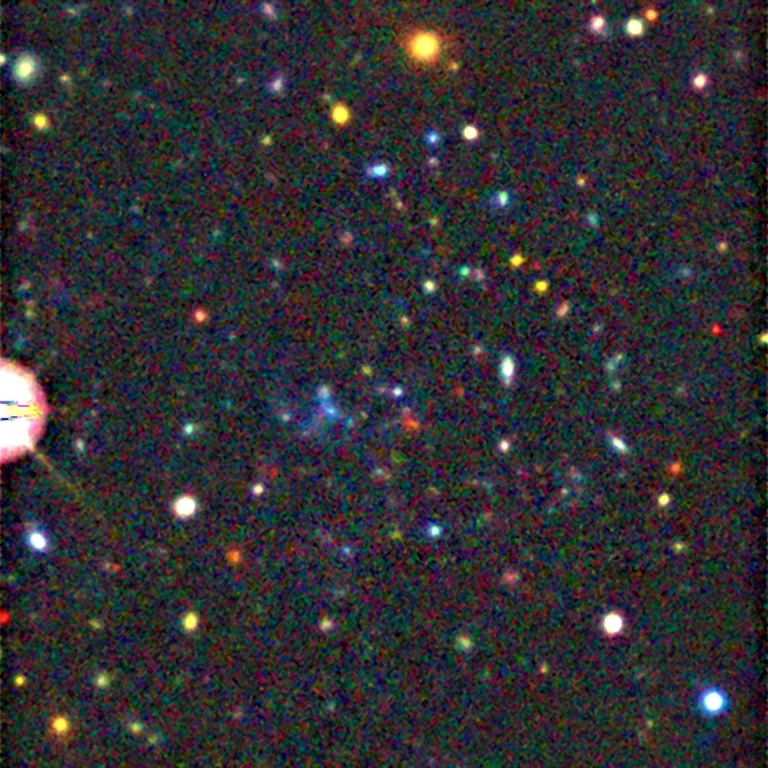}
\includegraphics[width=0.23\linewidth]{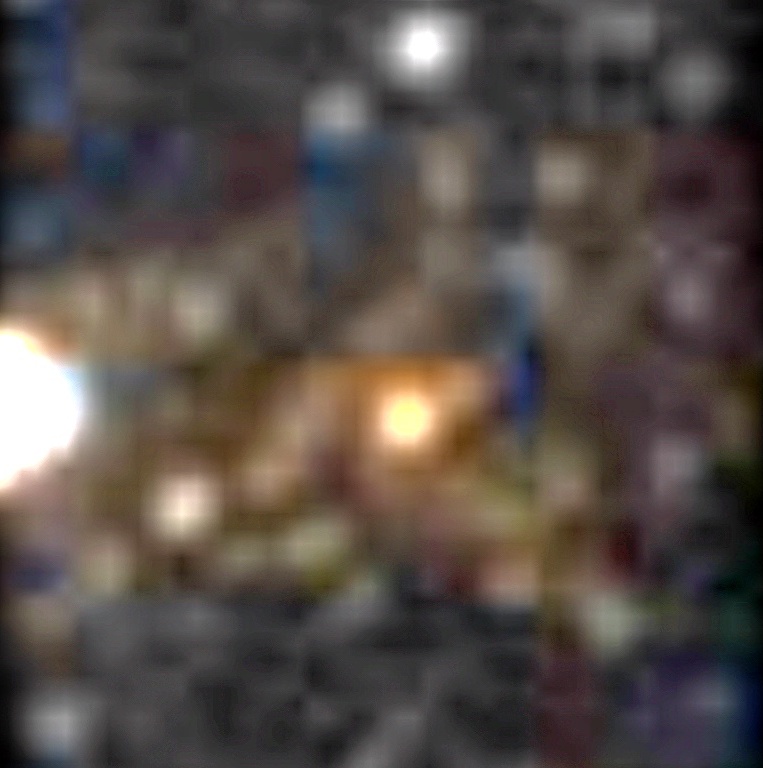}
\includegraphics[width=0.26\linewidth]{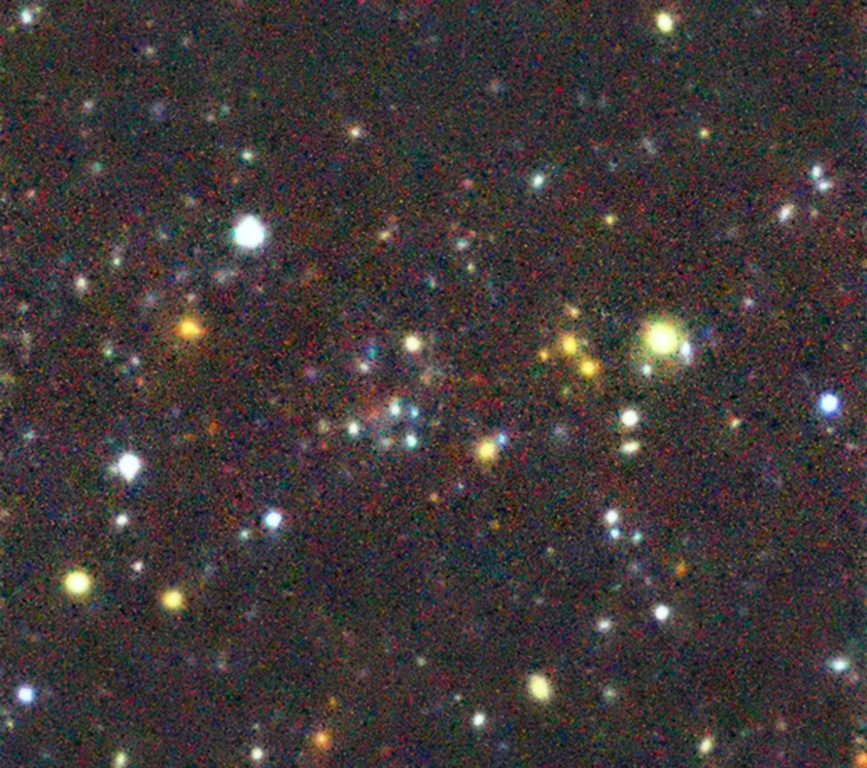}
\includegraphics[width=0.26\linewidth]{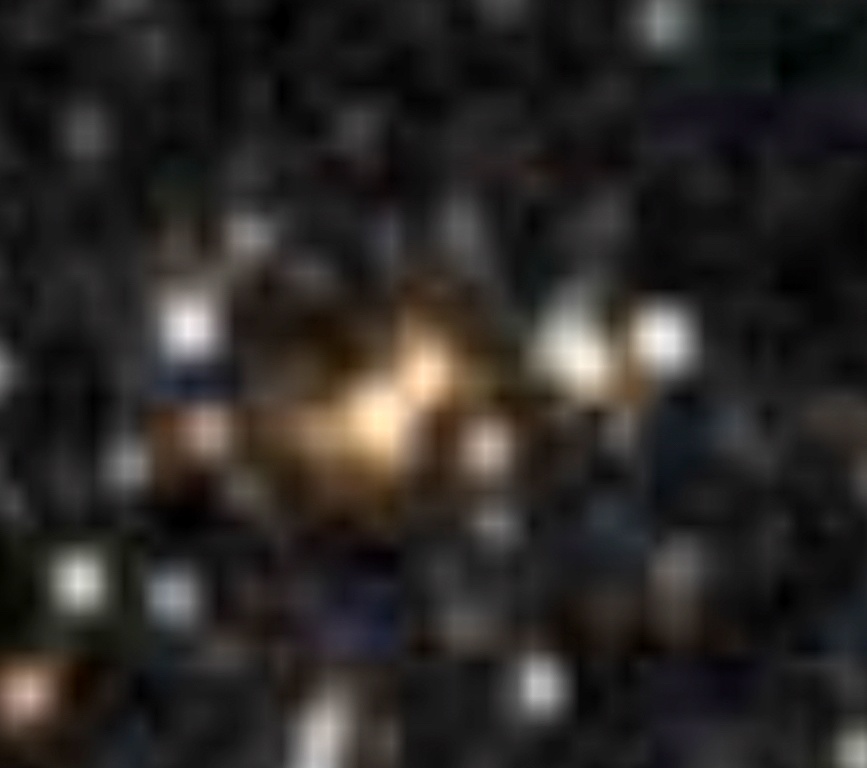}
\caption{High redshift galaxy clusters with ongoing star formation. The first two color composite images show ACT-CL J0431.0-3236 ($z=1.54$, SNR=5.1) in optical $g,r,z$-bands (left most image) and in WISE $w1,w2$-bands (second from left). The third and fourth images show the same optical and WISE images for ACT-CL J1240.4+0534 ($z=1,8$, SNR=5.3). While the BCGs are visible for both clusters in optical and WISE, we see an unusually high concentration of blue galaxies near the cluster core.}
\label{fig:HighZclusters}
\end{center}
\end{figure*}
The ACT-DR5~MCMF cluster catalog represents a major step forward in the number of high redshift ICM-selected galaxy clusters. Compared to the
predecessor catalog \citep{ACTDR5}, the new sample contains three times more clusters above $z=1$. In addition, it starts to reach redshifts close to the peak of cosmic star formation at $z=2$, a redshift where the red sequence of cluster galaxies is being established.
With $\sim$700 clusters at $z>1$, this sample contains the largest number of high redshift ICM-selected clusters to date. Current MCMF photo-z's suggest $\sim$110 to be at even higher redshifts of $z>1.4$. When looking at clusters at higher redshift, we tend to see more star forming galaxies in the clusters. Given the rest-frame sampling of the DECam optical bands, we start to miss passive galaxies in the LS-DR10 imaging data and see preferentially star forming galaxies. Only in combination with WISE NIR data can we also study the passive population in the high redshift regime. In Fig.~\ref{fig:HighZclusters}, we show two examples of high redshift clusters. The BCG is visible as a red galaxy in the optical and is also prominent in the WISE-based images as the brightest source in the center of the cluster. Aside from the BCG, few other red galaxies are visible, while both optical images show blue galaxies potentially associated with the clusters.

\begin{figure*}
\begin{center}
\includegraphics[width=0.315\linewidth]{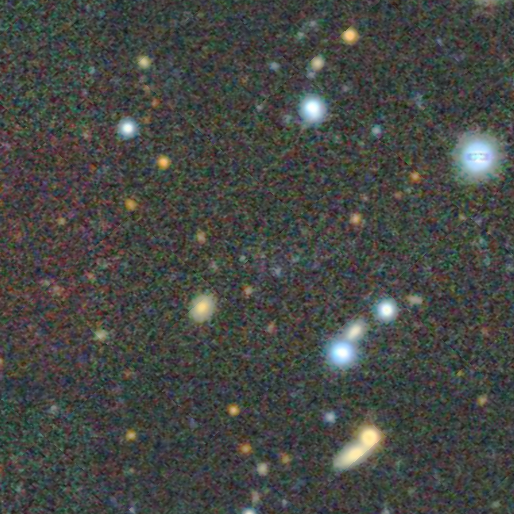}
\includegraphics[width=0.315\linewidth]{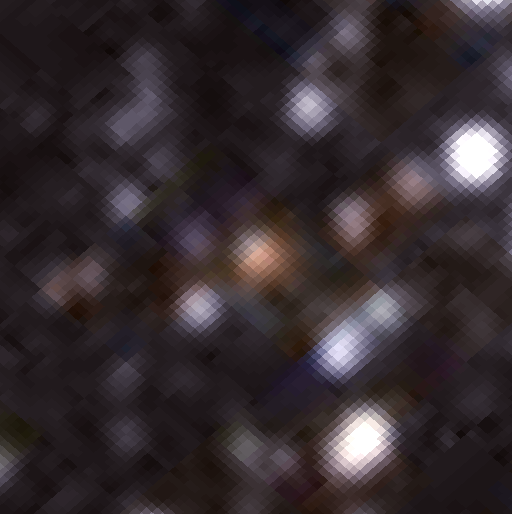}
\includegraphics[width=0.32\linewidth]{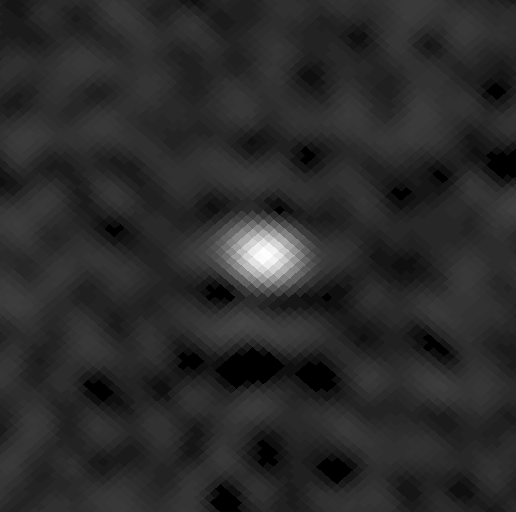}
\caption{ACT-CL J1424.7+0642, a SNR=7.35 ACT-DR5 candidate that is not confirmed with MCMF. Left: The LS-DR10 $g,r,z$-band color composite does not show an excess of red galaxies. Middle: WISE $w1, w2$-band color image, showing a bright WISE source at the location of the tSZE candidate. Right: 1.36 GHz image of the Rapid ASKAP Continuum Survey \citep{RACS}, showing a bright source at the WISE location, indicating the presence of a massive and passive galaxy.}
\label{fig:ACT-CLJ1424.7+0642}
\end{center}
\end{figure*}

We wish to stress that the MCMF confirmation using LS-DR10 and WISE is reaching its limits at high redshift and that some increased incompleteness is expected at the high redshift end because of that. This circumstance might be highlighted with the example of ACT-CL J1424.7+0642, a previously unknown cluster that is not confirmed as a member of the ACT-DR5~MCMF cluster sample. ACT-CL J1424.7+0642 at RA=216.191 DEC=6.715 has a SNR of 7.35, which makes it very unlikely (0.26\%) to be a noise fluctuation (see Figs.~\ref{fig:nfalsecandidates}~\&~\ref{fig:candidatelistpurity}). The optical LS-DR10 image, shown in Fig.~\ref{fig:ACT-CLJ1424.7+0642}, does not show any excess of red galaxies, and only a hint of some blue galaxies at the tSZE position. In the WISE image we find a bright source, with no optical counterpart close to the tSZE position and no evidence for additional cluster galaxies. Interestingly, this cluster has a bright Radio counterpart, as shown in the 1.36 GHz image of the Rapid ASKAP Continuum Survey \citep{RACS} in Fig.~\ref{fig:ACT-CLJ1424.7+0642}. The combination of a bright Radio and bright WISE source indicates the presence of a massive passive galaxy like the BCG of a galaxy cluster. The lack of optical emission at the same time suggests that this cluster must be at high redshift. MCMF estimates a tentative redshift of $z=1.59$ but given that the cluster fails MCMF confirmation (\fcont=0.5) this redshift is more indicative than definite.

Deeper imaging data and NIR wavelength coverage are needed to confirm and study these high redshift ACT-DR5~MCMF clusters. In this context, the ongoing survey by the Euclid mission \citep{Euclidsurveys} seems to be a particularly promising source of such data. However, such a sizeable and uniformly selected high redshift ensemble of clusters would also be well suited for dedicated, deep observations with other observatories.

\begin{figure}
\begin{center}
\includegraphics[width=0.99\linewidth]{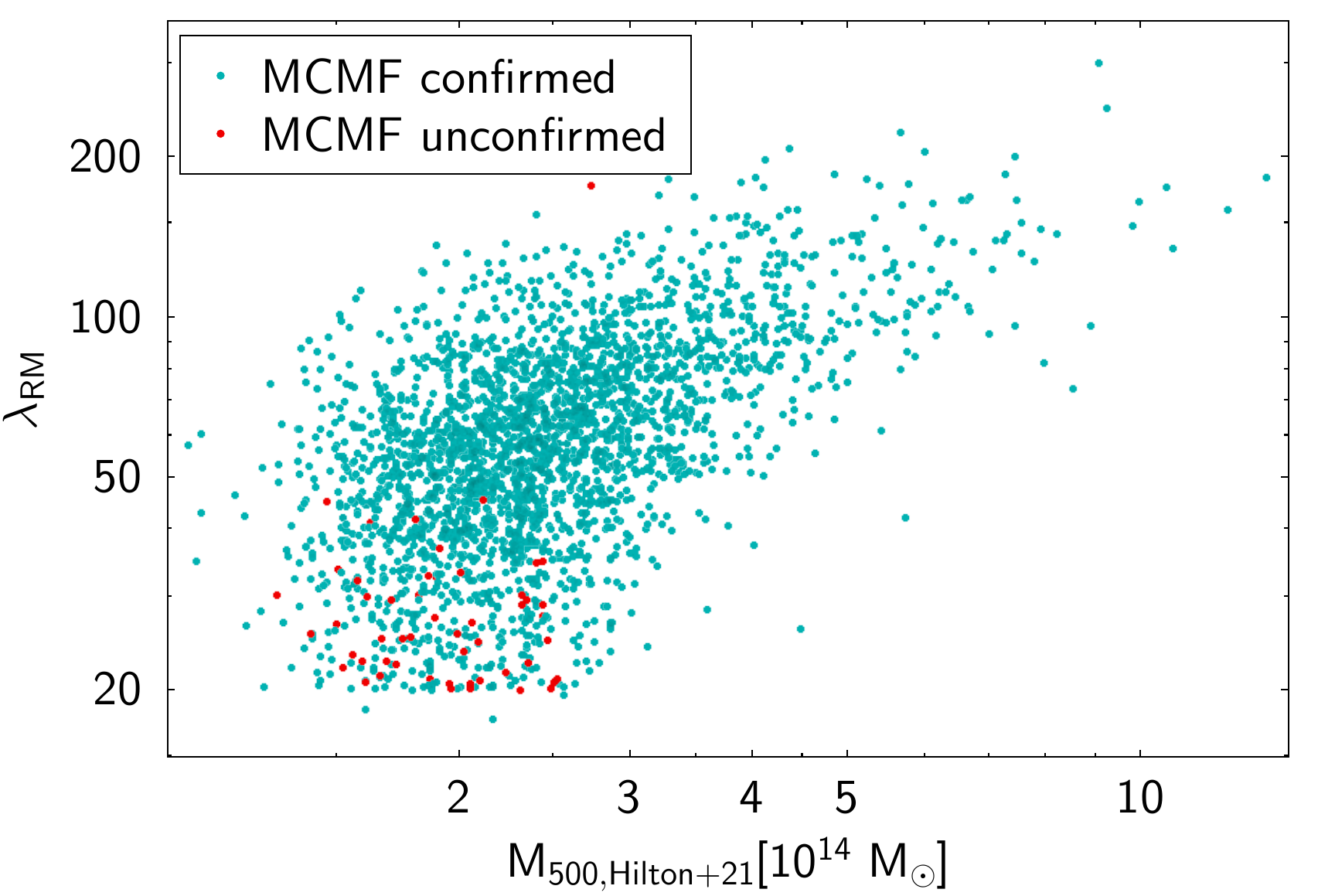}
\caption{Richness versus tSZE mass of clusters from the previous ACT-DR5 catalog \citep{ACTDR5}. Richness is either from DES or SDSS redMaPPer \citep{redmapper} or CAMIRA \citep{CAMIRA-HSC} and scaled to account for systematic uncertainties between redMaPPer and CAMIRA. Clusters also confirmed by MCMF are shown in blue, while those not confirmed with MCMF (red points) show up largely at low richness and tSZE-based mass.}
\label{fig:ACTxACTrichnessmass}
\end{center}
\end{figure}

\subsection{Comparison to the previous ACT-DR5 cluster catalog}
The previous ACT-DR5 cluster catalog \citep{ACTDR5} employed a mix of surveys and techniques to obtain redshifts and to confirm cluster candidates. The main method to obtain redshifts and confirm clusters was a positional cross-match between the candidate tSZE position and the positions of optically selected clusters. In the case of sources in the DES \citep{DES2016} and SDSS \citep{SDSS} regions, cross-matches were performed to optical clusters identified by the redMaPPer algorithm \citep{redmapper} with settings that restricted the fraction of chance matches to a level of 5\%. A similar approach was used over the HSC-SSP \citep{HSC-SSP} footprint, matching with optically selected clusters from the CAMIRA catalog \citep{CAMIRA-HSC}. Together, this cross-matching of tSZE candidates to optically selected cluster catalogs resulted in the confirmation of 1,491 clusters ($\sim35.5$\%). For the remaining 64.5\% of the previous ACT-DR5 cluster sample other less controlled methods or surveys were used. Although not explicitly stated, this suggests that the previous ACT-DR5 catalog can be expected to have a contamination of $\geq5$\%.

We adopt a maximum distance of 1.5\,arcmin to cross match our ACT-DR5 candidate list with the previous ACT-DR5 cluster catalog \citep{ACTDR5} and find 12 sources ($0.3$\%) from \citep{ACTDR5} not matched to our candidates. Most of these sources lie close to masked point sources in the ACT map. We therefore conclude that mild changes in the source detection pipeline and point source masking scheme introduces a very minor difference in our candidate list with respect to the version used in constructing the previous catalog \citep{ACTDR5}.

From the 4,182 matched sources we find 4,126 (98.7\%) to lie simultaneously in LS-DR10 and WISE unmasked footprints and 4,157 ($99.5$\%) to lie at least in one of the two footprints.
We re-confirm 4,007 ($\sim96.5$\%) of the previously published clusters \citep{ACTDR5} when adopting a \fcont<0.2 threshold, leaving 129 of the previous clusters ($\sim3.5$\%) unconfirmed even though they have decent LS-DR10 and WISE coverage.

In Fig.~\ref{fig:ACTxACTrichnessmass} we show the distribution of richness versus tSZE-based mass for the subset of the sample that have either redMaPPer or CAMIRA measurements available in the previous catalog \citep{ACTDR5}. We account for an offset between redMaPPer richness and CAMIRA richness by multiplying the CAMIRA richness by a factor of 1.5.
As visible from the plot, the majority of the sources not confirmed with MCMF lie at low richness and below the main scaling relation between richness and tSZE-based mass estimate provided in \citet{ACTDR5}. This is indicative that a significant fraction of the sources not meeting the MCMF selection threshold are indeed chance super positions instead of massive clusters. This is further supported by the fact that the fraction of clusters confirmed with redMaPPer and CAMIRA, for which \citet{ACTDR5} applied a systematic cut to allow only 5\% chance positions, drops from 63\% for MCMF confirmed to 40\% for the MCMF unconfirmed cases. It is worth noting that the cross-matched redMaPPer and CAMIRA clusters show visibly more scatter between richness and tSZE-based mass than the MCMF-based richness shown in the right plot of Fig.~\ref{fig:ACTredshift}.

Although a significant fraction of the systems unconfirmed with MCMF are indeed noise fluctuations, we do find evidence that some clusters are also unconfirmed. Those missed are either affected by small scale masking (due to, e.g., bright stars), are at high redshift (e.g., ACT-CL J0217.7-0345, $z=1.91$), where we hit the limitations of the optical and NIR imaging surveys or are at very low redshift (e.g., ACT-CL J2313.9-4243,$z=0.056$). Regarding the issues at low redshift, it is well known that tSZE detection from surveys at 90 and 150\,GHz becomes increasingly insensitive when clusters reach angular scales where the primary CMB fluctuations peak. This behavior is also reflected in the MCMF selection, where the richness threshold $\lambda_\mathrm{min}(z)$ increases toward low redshift. Because this increase is steep and may depend on the local survey properties, there are cases where even though the cluster is optically well detected, the cluster richness does not lie above the MCMF threshold.

Finally, there are 2,226 ACT-DR5~MCMF clusters (\fcont<0.2) that do not have a match in the previous catalog \citep{ACTDR5}. A certain fraction of these clusters simply were not in the candidate list used in \cite{ACTDR5}, because in our current analysis we lower the detection threshold.
In addition, the authors of the previous study estimated that $\sim 960$ clusters were missing from their sample.  Together then it is not surprising that we have identified more clusters in the ACT-DR5~MCMF catalog. 

The SNR>5 regime was of special interest in the previous ACT-DR5 catalog \citep{ACTDR5} and therefore deserves specific mention.  Candidate clusters in this SNR regime were visually inspected in the previous analysis to ensure their quality. Furthermore, we expect that almost all SNR>5 sources that appear in our candidate list were also present in the previous sample, although potentially at slightly different SNR values. It is therefore interesting that we confirm 327 new clusters at SNR>5, and we fail to confirm 18 clusters despite good imaging data at those locations. These 327 new clusters correspond to $\sim$12\% of all clusters with SNR>5 and therefore it indicates a significant incompleteness in the previous ACT-DR5 catalog \citep{ACTDR5}.

\section{Comparison of ACT-DR5~MCMF to  other catalogs}\label{sec:crossmatches}

Given the location and size of the ACT-DR5 footprint, the number of cluster catalogs with possible overlap with the ACT-DR5~MCMF catalog is too large to discuss them all.
We therefore restrict the comparison to the largest ICM-selected cluster samples. 
When comparing to other catalogs, we typically want to extract the total number of clusters in common as well as the list of concrete matched clusters. We are also interested in sources that are not matched 
but should have been, given their mass and redshift.

To get an estimate of the total number of matched clusters, we first cross match the samples using a large matching radius of $0.2^\circ$. We repeat the matching procedure after shifting the source positions by $\pm2^\circ$ to estimate the number of chance matches.  Using shifted source positions allows us to include the impact of depth variations with position in the ACT-DR5 survey. Using the offset distribution of the matched sources for the normal and the shifted catalog one can measure the number of true matches as a function of positional offset under the assumption that both offset distributions are composed of chance matches at large offsets ($\Theta\approx0.2^\circ$).
The total number of matches is simply the integral over the offset distribution minus the integral over the scaled offset distribution of the chance matches. We can use the same distributions to identify a matching radius that yields a reasonable compromise between completeness and contamination by chance matches. 

\subsection{SPT candidate and cluster catalogs}
The SPT collaboration has also performed multiple mm-wave surveys, with a series of receiver upgrades over the years.  Most noteworthy are the SPT-SZ survey \citep{Bleem15}, the SPT-ECS survey \citep{Bleem20} and the recently published SPTpol~500d survey \citep{Bleem24}. The MCMF informed cluster catalog from SPTpol~500d, a new version of the SPT-SZ cluster catalog \citep[][SPT-SZ MCMF]{Klein24a} and an MCMF augmented version of the SPT-ECS cluster catalog were then used for the cosmological study presented in \citet{Bocquet2024arXiv240102075B}.

Of special interest for this work is the SPTpol~500d survey, because it is on average significantly deeper than the ACT-DR5 dataset and therefore it allows one to test the purity and completeness of the ACT-DR5~MCMF cluster catalog. To test for completeness we first match all ACT-DR5 candidates over the SPTpol~500d footprint with the SPTpol~500d candidate list.
Given the reasonably small positional uncertainties and low source densities, we find that a cross-match radius of 2\,arcmin includes all true cluster matches, while only one chance match is expected. We find 118 of the 339 ACT-DR5 cluster candidates over the SPTpol~500d field have matches in the SPTpol~500d candidate catalog.
From those 118 matches, 115 (97\%) are confirmed in our ACT-DR5~MCMF sample (\fcont<0.2). Of the three missing systems, only one (ACT-CL J2247.6-5537) is a confirmed cluster in SPTpol~500d, and it suffers from masking in the optical/NIR survey data due to a nearby star. From the remaining two clusters, one (ACT-CL J2342.9-5521) shows the largest SPT to ACT positional offset (99 arcsec) out of all matches. Despite the large offset and potential small scale masking impeding confirmation, we assume at least two of the three to be true clusters. Given the number of matches this yields an estimate of the completeness of the ACT-DR5~MCMF catalog of $\sim$98\%, close to the expected completeness of 97.8\% presented in Table~\ref{tab:sample}.

Out of the 221 ACT-DR5 candidates in the common footprint without a match we find that 52 are members of the ACT-DR5~MCMF cluster catalog (\fcont<0.2). Out of those we find ten that are eRASS1 extent selected clusters and an additional seven that are eRASS1 point source (see discussion in Sec.~\ref{sec:erass}); using optical, NIR, eRASS1 and ACT-DR5 maps, we identify another 19 candidates that are likely real clusters. Thus, 36 out of the 52 ACT-DR5~MCMF clusters (\fcont<0.2) without SPTpol~500d counterparts are likely real clusters. In total this yields 151 confirmed real clusters in the SPTpol~500d footprint, out of 167 clusters in the ACT-DR5~MCMF catalog (\fcont<0.2). This corresponds to a sample purity of 90.4\%, close to the 
expected 89.7\% presented in Table~\ref{tab:sample}.  With 154 real clusters out of 339 candidates in the the SPTpol~500d footprint, we can estimate a contamination level of the ACT-DR5 candidate list to be $\sim$55\%, close to the value estimated in Sec.~\ref{sec:initialconta} of $53.7\pm0.5$\%.

The SPT-SZ MCMF and SPT-ECS catalogs are shallower than SPTpol~500d and therefore do not allow us to test completeness and purity to the same extent as SPTpol~500d. On the other hand, their large footprints of $\sim$2,500 deg$^2$ provide more clusters in common that could be used for cross-calibration studies.  We find that using the cross-match radius of 2\,arcmin yields 435 matches between SPT-SZ MCMF confirmed clusters and the ACT-DR5 cluster candidates, from which 431 (99\%) are members of the ACT-DR5~MCMF cluster catalog (\fcont<0.2).  Repeating the same exercise with SPT-ECS clusters, we find 290 matches with the ACT-DR5 candidate list, of which 288 (99.3\%) are members of the ACT-DR5~MCMF cluster catalog. Taking the overlap between the different SPT-based catalogs into account, we find a total of 768 matches between those three SPT cluster catalogs and our ACT-DR5~MCMF cluster catalog. This sample is sufficiently large to allow for cross checks of, for example, the tSZE inferred mass or the survey selection function models.

In summary, the cross-comparison with the deeper SPTpol~500d survey confirms the completeness and contamination levels we present for the ACT-DR5~MCMF catalog as well as our measurement of the initial contamination level of the ACT-DR5 candidate list.

\subsection{RASS-MCMF cluster catalog}\label{sec:ACTxRASS}
The RASS-MCMF cluster catalog \citep{Klein23} covers 25,000\,deg$^2$ of extragalactic sky and contains 8,449 clusters. The optical data used for cluster confirmation is identical to that used in the current analysis, and we expect only minor departures from complete coverage of our ACT-DR5 footprint due to RASS low exposure time regions.
We find 1,061 of the RASS-MCMF clusters in our ACT-DR5~MCMF catalog.  The offset distribution of matches shows an excess over chance matches out to $\sim$5.5\,arcmin.  Using a matching radius of 4\,arcmin yields 1,066 matches with an expected contamination by chance matches of 1.7\% and an incompleteness (i.e., true matches with offsets >4\,arcmin) of 1.3\%. This number of matches corresponds to $\sim$16\% of the full ACT-DR5~MCMF sample. Conversely, $\sim$29\% of the RASS-MCMF sample within the ACT-DR5 footprint is also present in the ACT-DR5~MCMF cluster catalog. 

The fraction of matched clusters is a strong function of redshift, mainly due to the different redshift sensitivity of the candidate selection in the X-ray and the tSZE. The fraction of RASS-MCMF clusters with an ACT-DR5~MCMF match rises steeply with redshift and flattens out at z>0.5, reaching fractions of $\sim$80-85\%. This match fraction is close to the expected purity of the RASS-MCMF sample of 90\%. We do not expect to match the full 90\% of the RASS-MCMF sample that consists of real clusters, because there is significant scatter in the X-ray and tSZE observable-mass relations and there are variations of the survey depths over the sky in both surveys.  Considering those effects, one can take the 80\% match fraction for clusters at z>0.5 to be a lower limit for the sample purity of the RASS-MCMF catalog and therefore a confirmation of the expected purity.

As mentioned in Section~\ref{sec:MCMFonLSDR10}, there are two important changes in the MCMF algorithm for this analysis that impact the cluster richness estimates.  Namely, we exclude the $w1-w2$ color from our MCMF richness estimate, and we improve the red sequence model for the $z-w1$ color. These changes weaken the rather strong redshift trends in the richness-mass relation seen in RASS-MCMF clusters, which are caused by the v-shaped redshift dependency of the $w1-w2$ color and the uncertainty in the redshift variation of the scatter in the $z-w1$ color (e.g. see Fig.1 in \cite{Klein24a}).  Because the richness algorithms for RASS-MCMF and ACT-DR5~MCMF differ, the resulting richness measurements for the same cluster appearing in the two catalogs also differ. We performed richness measurements using both richness algorithms for RASS-MCMF and ACT-DR5~MCMF clusters to model the differences in richness, and we provide a transformation (see Table~\ref{tab:oldnewrich}) that can be used to calculate updated richnesses for the RASS-MCMF clusters that are equivalent to those one would obtain using the current MCMF settings. 

We have examined the richnesses coming from the previous and current MCMF algorithms, finding that their scatter about the ICM based mass estimates is similar.  That is, both types of richness estimates are valid, internally self-consistent, and can be used as mass proxies with similar scatter. The remaining differences between RASS-MCMF and ACT-DR5~MCMF richnesses are then just the consequence of using different ICM-based priors on cluster center and mass.

\subsection{eRASS1 cluster catalog}\label{sec:erass}
\begin{figure}
\begin{center}
\includegraphics[width=0.99\linewidth]{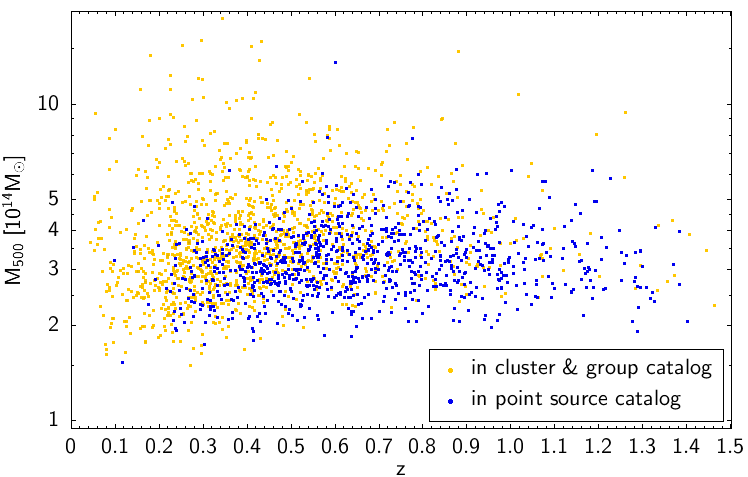}
\caption{eRASS1 matched ACT-DR5~MCMF clusters in mass versus redshift. Matches to the extent selected cluster and group catalog are highlighted in yellow.   ACT-DR5~MCMF clusters matched to sources in the eRASS1 point source sample are shown in blue.  One famous ACT-DR5 cluster missed in the eRASS1 cluster sample is ACT-CL J2344.7-4243 (the Phoenix cluster).} \label{fig:eRASSxACTmz}
\end{center}
\end{figure}

\begin{figure}
\begin{center}
\includegraphics[width=0.99\linewidth]{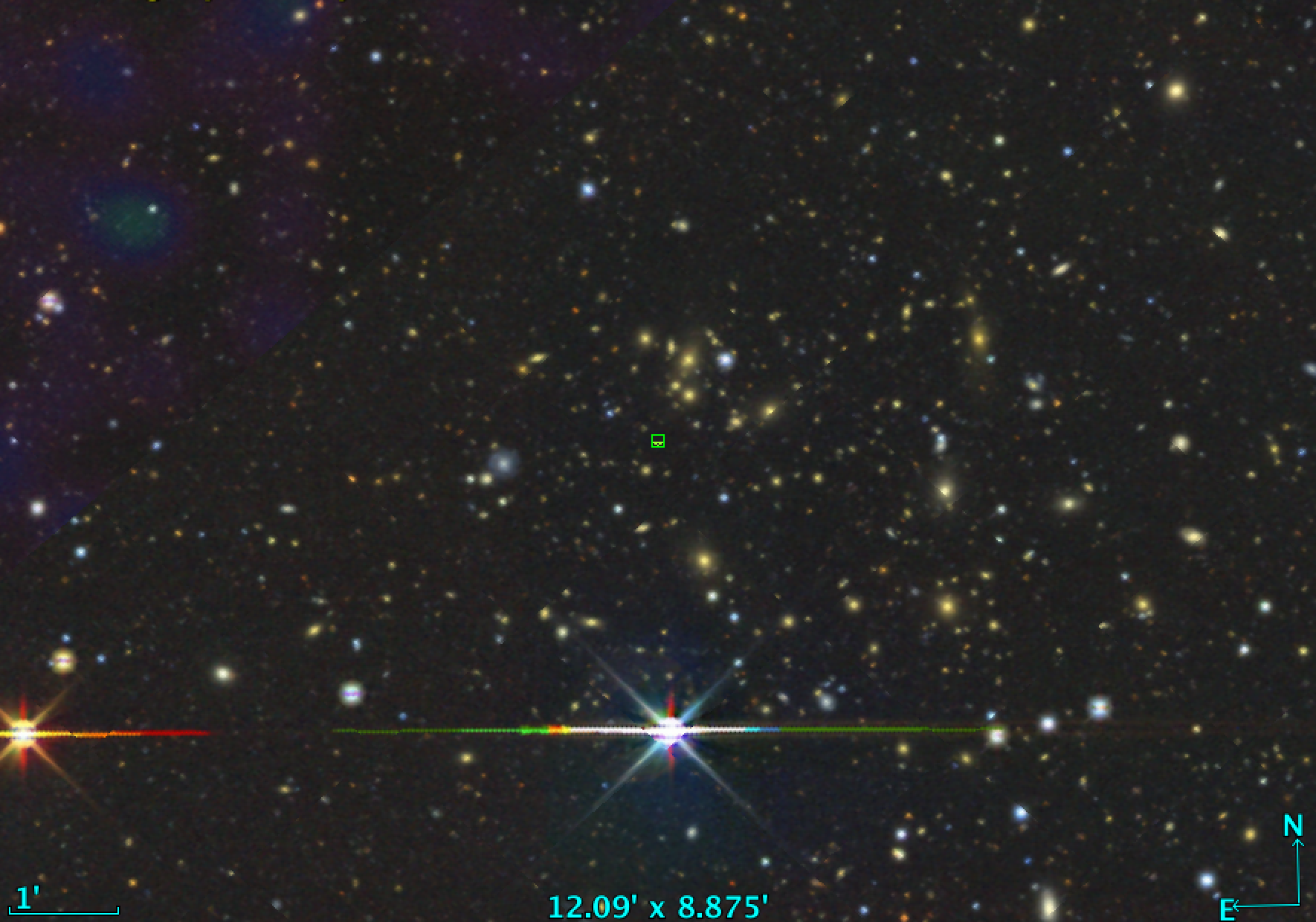}
\includegraphics[width=0.99\linewidth]{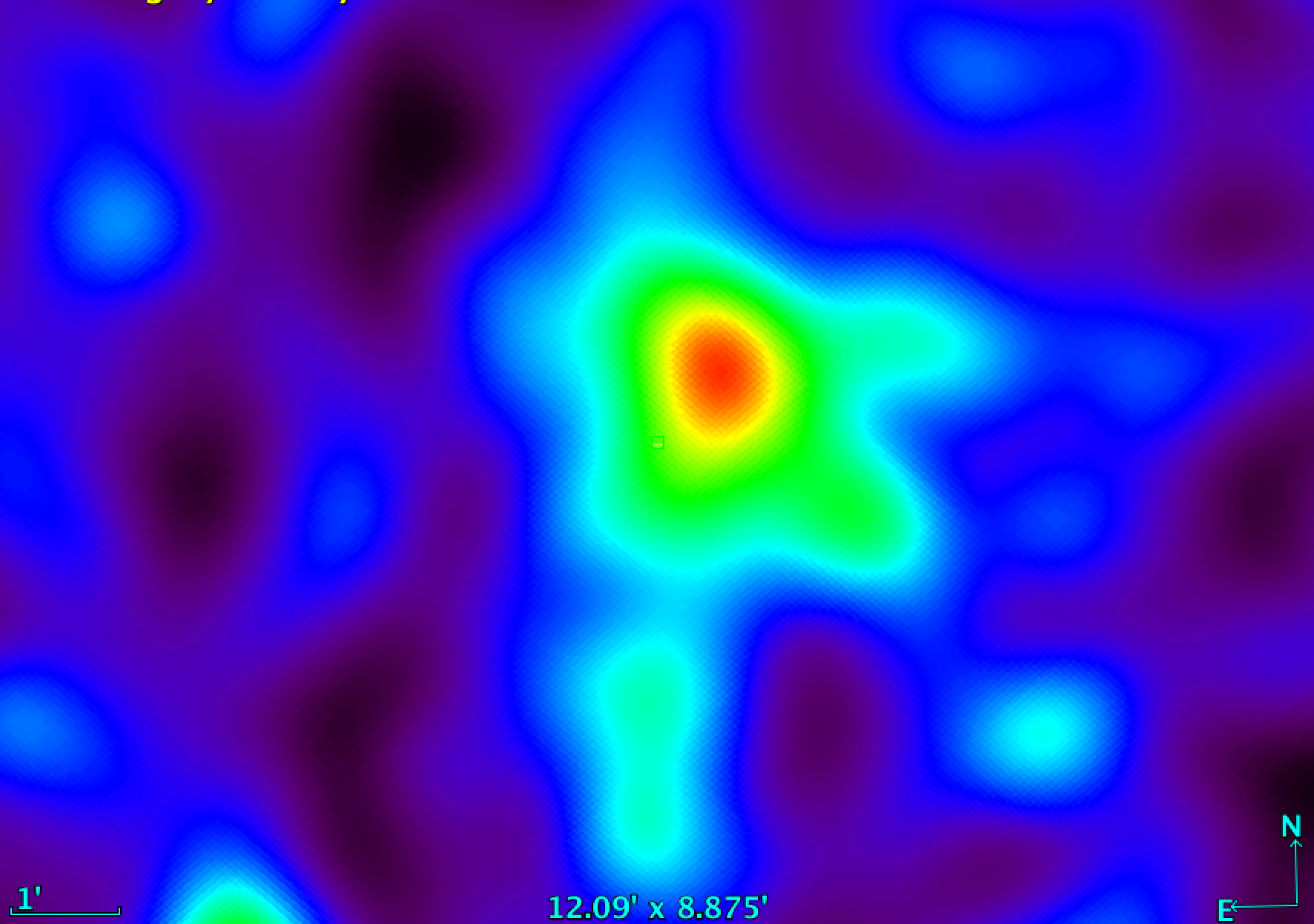}
\caption{Optical LS-DR10 image (top) and smoothed eROSITA count rate image (bottom) of ACT-CL J1416.5+0534 (z=0.23, SNR=4.9) provide one example of a low redshift ACT-DR5~MCMF cluster only matched to the eRASS1 point source catalog.} \label{fig:ACT-CLJ1416.5+0534}
\end{center}
\end{figure}
\begin{figure}
\begin{center}
\includegraphics[width=0.99\linewidth]{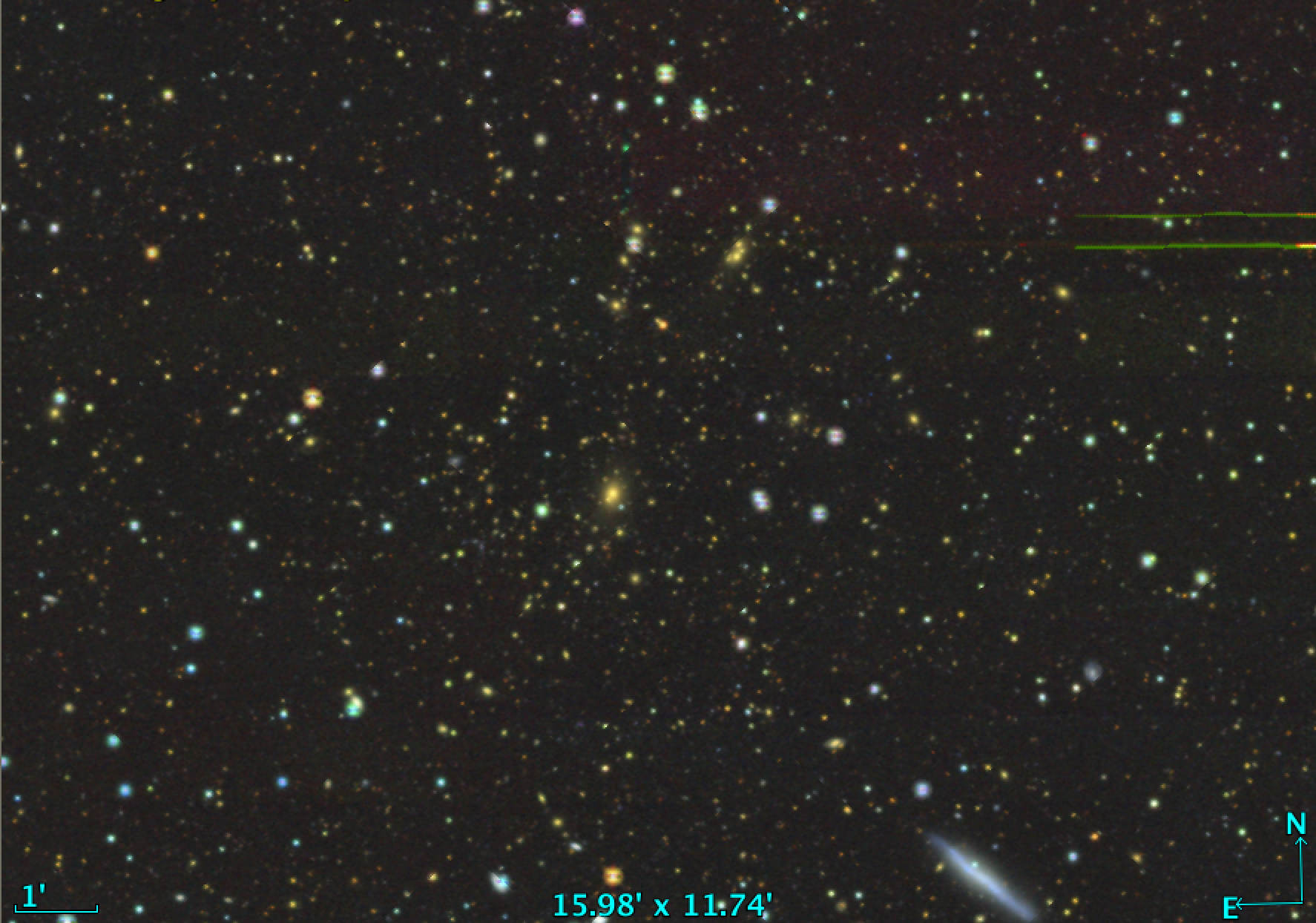}
\includegraphics[width=0.99\linewidth]{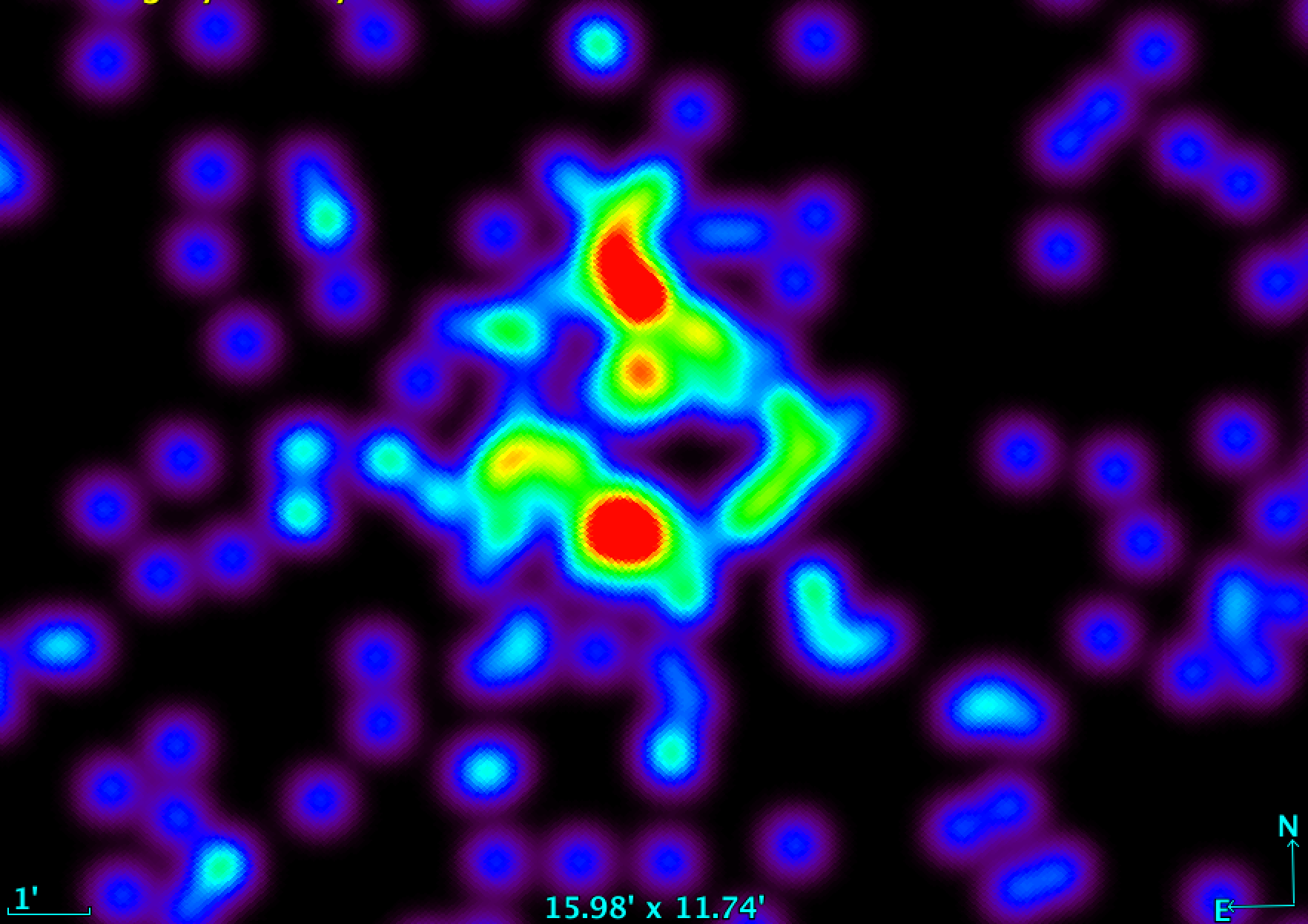}
\caption{Optical LS-DR10 image (top) and smoothed eROSITA count rate image (bottom) of ACT-CLJ0840.5+0543 (z=0.28, SNR=9.75). This is one of 924 ACT-DR5~MCMF clusters with matches in the eRASS point source catalog. This particular cluster, along with $\sim$70 other clusters in the common eRASS1-ACT-DR5 footprint, is missing from the eRASS1 cluster sample but is included in the RASS-MCMF cluster sample.} \label{fig:ACT-CLJ0840.5+0543}
\end{center}
\end{figure}
The eROSITA collaboration recently released the first survey data covering the western Galactic hemisphere \citep{Merloni24}. In addition to the catalog of all detected X-ray sources, there is an extent selected cluster and group catalog \citep{Bulbul24}. The cluster and group catalog is of special interest, because it serves as the basis of the eROSITA eRASS1 cosmology analysis \citep{eRASS1cosmo}, in which the eROSITA team reports an unusually high value of $S_8$ compared to the majority of other recent cluster and large scale structure analyses 
\citep[e.g.][]{WtG,KidsClusterCosmo,Chiu23,Bocquet2024arXiv240102075B,DESpKIDS,ACTDR6lensing}.

We first match the ACT-DR5~MCMF sample to the cluster and group catalog. We find 1,306 matches, almost all (>99\%) showing offsets smaller than 3.5\,arcmin. Imposing a maximum offset of 3.5\,arcmin yields 1,326 matches with an expectation of $\sim$20 chance matches, corresponding to a contamination level of 1.5\%.  We then remove matches to the eRASS1 cluster and group catalog and match the remaining ACT-DR5~MCMF clusters to the eRASS1 point source catalog, obtaining 965 matches in excess to the expected number of chance matches.  In total then, of the 2,271 ACT-DR5~MCMF clusters that match to eRASS1 sources, 43\% are classified as point sources rather than clusters and groups.

In Fig.~\ref{fig:eRASSxACTmz} we show the distribution of eRASS1 matches in mass versus redshift. The eRASS1 X-ray extent and extent likelihood selected sample of clusters and groups results in a dramatic loss of clusters compared to the sample that matches both X-ray point source and extended source catalogs. This impacts clusters with masses $M_{500}\approx2\times10^{14}$ M$_\odot$ at intermediate redshifts z$\sim$0.35 and beyond. The X-ray extent selection suffers from incompleteness when there is strong emission from the cluster center. One clear example of this is one of the most massive clusters missed in the eRASS1 cluster and group catalog, which is ACT-CL J2344.7-4243 (the Phoenix cluster), which is visible in Fig.~\ref{fig:eRASSxACTmz} at $z\approx0.6$ and $M_{500}\approx10^{15}$ M$_\odot$. 

Examining some of the low-z ACT-DR5~MCMF clusters that appear in the eRASS1 point source sample, we find two additional reasons for clusters to be lost from the eRASS1 cluster and group catalog: source splitting and low surface brightness. In Fig.~\ref{fig:ACT-CLJ1416.5+0534} we show the region around ACT-CL J1416.5+0534 ($z=0.23$, SNR=4.9, $M_{500}=2.1\times10^{14}$ M$_\odot$). The smoothed eRASS1 map shows clear evidence for extended emission at the cluster location, but this eRASS1 source is classified as point like. Given the clearly apparent extent of the cluster, the most likely reason for this classification is that the extent likelihood is low. 

In Fig.~\ref{fig:ACT-CLJ0840.5+0543} we show another case of an ACT-DR5~MCMF cluster missed in the eRASS1 cluster and group catalog: ACT-CLJ0840.5+0543 (z=0.28, SNR=9.75, $M_{500}=4.8\times10^{14}$ M$_\odot$). The X-ray emission of the cluster is apparent, but it seems to be distributed in different patches and potentially impacted by a real point source in the northern clump. This cluster is in fact bright enough to be one out of $\sim$70 ACT-DR5~MCMF clusters that are listed as eRASS1 point sources and simultaneously are confirmed RASS-MCMF clusters \citep{Klein23}.

Despite the fact that we identify numerous cases of clusters only appearing in the point source sample, it is not immediately clear if this could lead to a bias in the estimated $S_8$ parameter found in \citet{eRASS1cosmo}. In fact the large amount of intrinsic scatter between the X-ray mass-proxy observable and the cluster mass found in \citet{eRASS1cosmo} results in a very broad selection function in cluster mass, allowing for clusters being missed (or perhaps more accurately-- mis-classified) even at very high masses. For example the cluster ACT-CLJ0840.5+0543 at $z=0.28$ has an eRASS1 exposure time of $150$ seconds, yielding only a 40\% detection probability even with a mass of $M_{500}=4.8\times10^{14}$ M$_\odot$ \citep[see Figure D.1 in][]{eRASS1cosmo}. What can be concluded from the cross matching is that a significant number of clusters are appear only in the eRASS1 point source sample, and that this is happening even at lower redshift and at masses greater than $M_{500}\approx2\times10^{14}$ M$_\odot$. 

It is worth noting that the eRASS1 cluster cosmological analysis \citep{eRASS1cosmo} is the first ICM-based cluster study that consciously adopts a significantly contaminated cluster sample ($\sim$15\% of the systems are not expected to be real clusters) and that attempts to account for the impact of non-cluster contaminants. An important ingredient in the so-called mixture model of real clusters and contaminants is the richness distribution along random lines of sight and along lines of sight towards eRASS1 point sources. The latter is meant to represent the richness distribution of AGN and other point like sources contaminating the extent selected cluster and group catalog. As we have shown here, there are in fact a large number of massive clusters in the point source catalog, which will populate the high richness regime in the richness distribution along lines of sight toward eRASS1 point sources.  That in turn biases the mixture model.

This is supported by the eRASS1 simulations \citep{eRASSsim} that indicate that $\sim$4.3\% of all eRASS1 sources are galaxy clusters.  Given the $\sim$930,000 detections in eRASS1 \citep{Merloni24} this translates into $\sim$40,000 clusters in eRASS1 from which only one quarter ($\sim$10,000) are found in the cluster and group catalog \citep{Bulbul24}. The richness distribution of point sources therefore contains three times more clusters than the actual eRASS1 cluster and group catalog. This may bias the richness distribution of the AGN model \citep{eRASSopt}, making it appear quite different from the richness distribution along random lines of sight.   We note that in the RASS-MCMF catalog a similar analysis showed no difference in the richness distribution along random lines of sight and the distribution along lines of sight toward AGN \citep{Klein23}.  Estimating the impact of this potentially biased modeling of the contamination on the   cosmological results \citep{eRASS1cosmo} requires further attention, but is 
beyond the scope of this paper.

\begin{figure*}
\begin{center}
\includegraphics[width=0.295\linewidth]{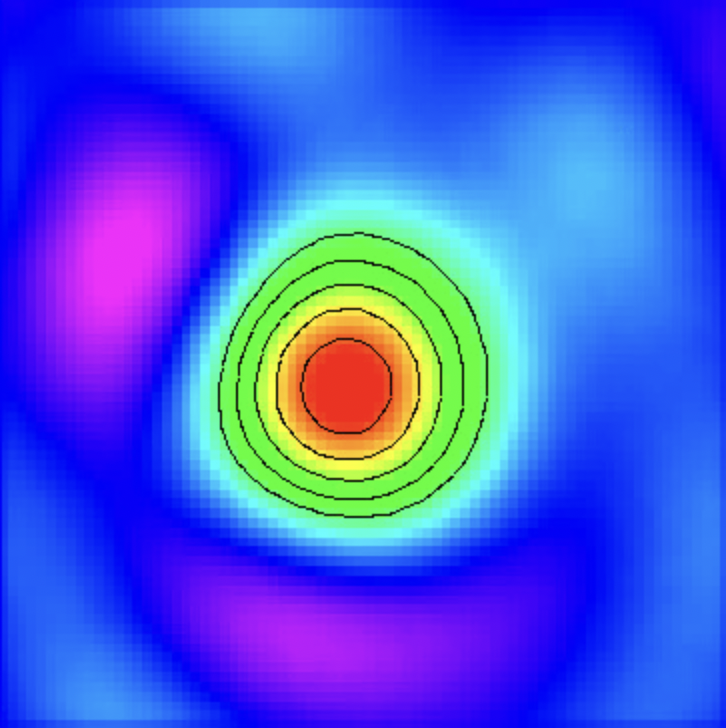}
\includegraphics[width=0.29\linewidth]{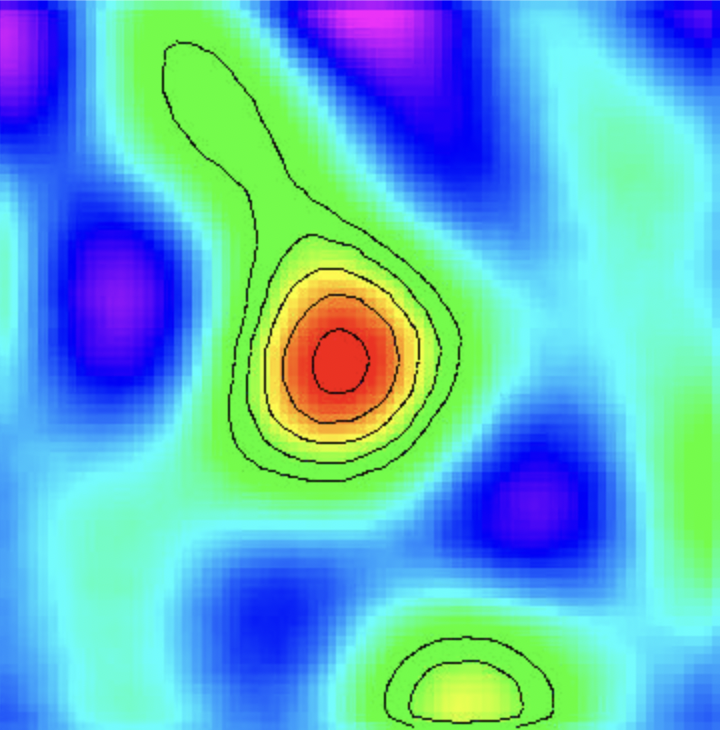}
\includegraphics[width=0.38\linewidth]{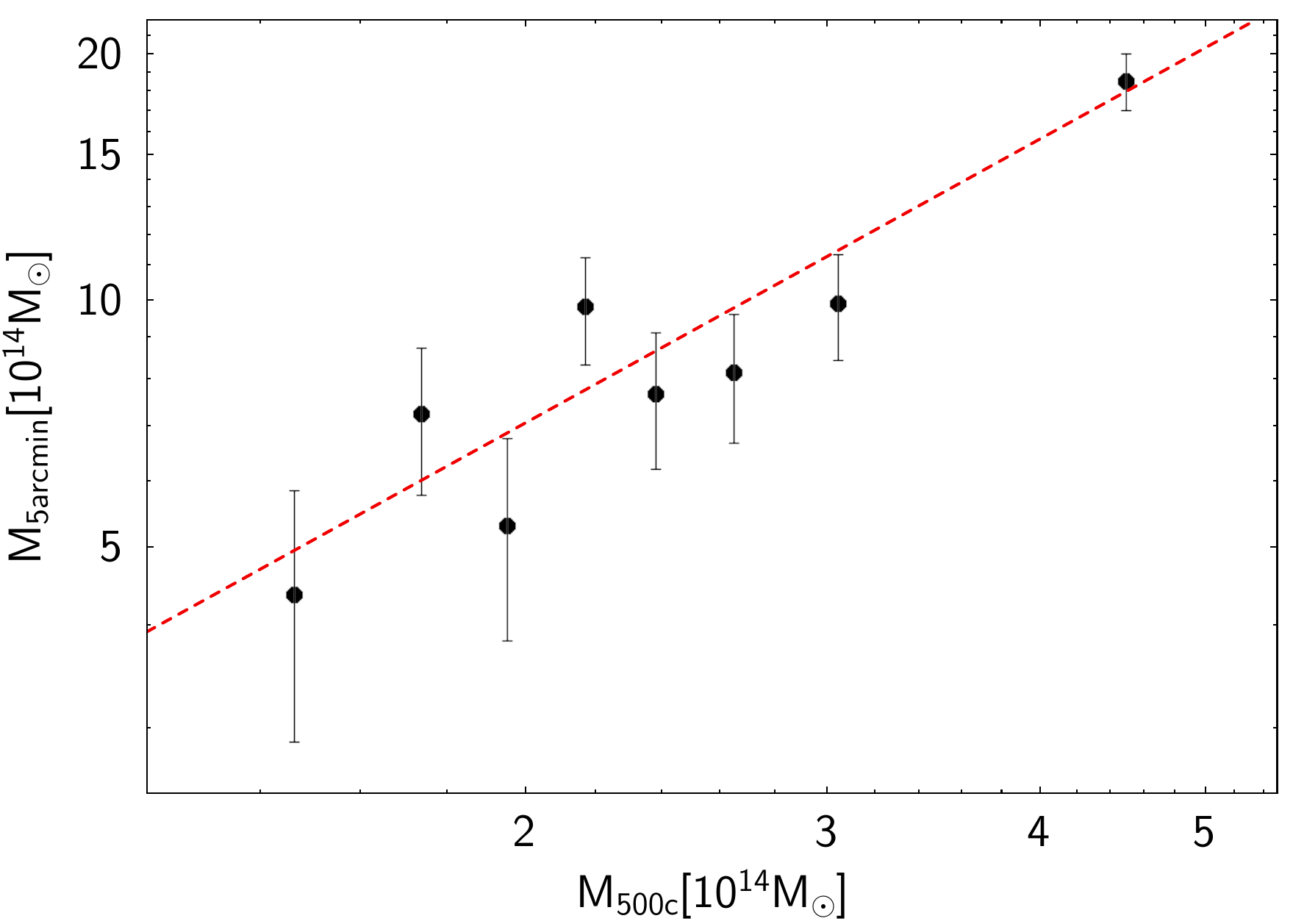}
\caption{Left \& middle: Stacked ACT-DR6 CMB lensing map for the full ACT-DR5~MCMF catalog (left) and and for the $z>1$ subsample (middle). Black contours show lensing convergence starting at $\kappa=0.03$ in steps of 0.01. The field of view is $25\times25$\,arcmin. Right: Projected CMB lensing mass within 5\,arcmin radius  $M_\mathrm{5\,arcmin}$ in bins of tSZE derived mass. The red line shows the best fit power law, suggesting a slope of $1.15\pm0.15$.} \label{fig:ACTlensing}
\end{center}
\end{figure*}

\section{Cross-correlation with ACT-DR6 CMB lensing}
We utilize the recently published ACT-DR6 CMB lensing map, \citep{ACTDR6lensing} which has $\sim$68\% overlap with our ACT-DR5~MCMF sample, to measure the CMB lensing signal associated with the MCMF confirmed cluster sample. Notable here is that the difference in footprints reflect a more restrictive masking applied for the CMB lensing study.

The CMB lensing map was constructed from the spherical harmonic $a_\mathrm{l,m}$ convergence coefficients, where the recommended multipole range for the maps is $600<l<2100$, resulting in an angular resolution of $\sim$5\,arcmin \citep{ACTDR6lensing}. The lensing map contains the convergence $\kappa=\Sigma/\Sigma_\mathrm{crit}$ along each line of sight.  
We use the ACT CMB lensing convergence maps to explore the significance of the detection of the ACT-DR5~MCMF sample.  Using the Purity-95 sample (see Tab.~\ref{tab:sample}), which provides a nice match of completeness and purity,
we show the mean CMB lensing convergence maps of the full sample (Fig.~\ref{fig:ACTlensing} left panel) and for the $z>1$ cluster subsample (right panel of same figure). The field of view shown is $25\times25$\,arcmin, corresponding to $\sim$5$\times$5 resolution elements in the $\kappa$ maps.  We obtain a clear, positive mean convergence signal for both cluster samples with signal-to-noise ratios of 16.4 and 4.3, respectively, within an aperture of 5\,arcmin. The estimated signal-to-noise ratios are based on measurements of 10,000 representative mock samples where clusters are positioned randomly within the CMB lensing convergence maps. We find that the values for the full ACT-DR5~MCMF cluster sample are consistent with the Purity-95 subsample.

The detection significance of 16.4\,$\sigma$ for the full ACT-DR5~MCMF cluster catalogue is the strongest CMB lensing signal found to date for a cluster sample. Additionally, the significance as well as the mean redshift of the high-z subsample (<z>=1.2) exceeds that of the optical+IR selected MadCows sample \citep{ActlensingMadcows} and currently represents the highest signal to noise high-redshift CMB lensing measurement for clusters. The strong detection of lensing convergence around the high redshift subsample highlights the fact that this ACT-DR5~MCMF sample includes many massive, high-redshift halos, in accordance with expectation.

For lines of sight containing a massive cluster, the weak gravitational lensing convergence is typically dominated by the surface mass density $\Sigma$ of the cluster.  Because $\kappa$ is the ratio of the cluster surface mass density to the critical surface mass density $\Sigma_\mathrm{crit}$, there is also a dependence on the cluster and source redshift.  Because the surface density of a galaxy cluster depends on the line of sight through the cluster, which scales as cluster mass $M_{500}^{1/3}$, the weak gravitational lensing convergence (or shear) provides mass information about the cluster that has long been used to calibrate cluster masses in ICM selected samples \citep[e.g.,][Singh et al, in prep]{WtG3,Hoekstra15,Dietrich2019MNRAS.483.2871D,Klein19lensing,Bocquet2019ApJ...878...55B, Bocquet2023arXiv231012213B,Grandis2024arXiv240208455G}.  

We therefore use the ACT-DR6 lensing convergence maps to explore the mass information that is present for the ACT-DR5~MCMF sample.  Namely, we divide the Purity-95 sample (see Tab.~\ref{tab:sample}) into mass bins, using the mass estimates following the original ACT-DR5 cluster catalog paper \citep{ACTDR5}. 
Similar to the previous analysis step, we create mean CMB lensing convergence maps but now for each mass bin.
Using the mean $\Sigma_\mathrm{crit}$ of the clusters contributing to a given mass bin, we calculating $\sum_i \kappa_i * \Sigma_\mathrm{crit}$ on the mean CMB convergence map for all pixels $i$ within 5\,arcmin from the cluster center.  We call the resulting mass estimate $M_\mathrm{5arcmin}$, corresponding to the cluster mass projected within a 5\,arcmin cyclinder around the cluster center. 
Within each mass bin the cluster sample spans a broad range of redshift, but at the median redshift of the sample the 5\,arcmin angular aperture corresponds to a physical radius of 2\,Mpc. Thus, the $M_\mathrm{5\,arcmin}$ notionally corresponds to the mass of the cluster within a cylinder with a radius of 2\,Mpc. This 2\,Mpc is close to the virial radius of the clusters in ACT-DR5~MCMF, and so the $M_\mathrm{5\,arcmin}$ can be considered to be a rough estimate of the total cluster mass.

In the right-most panel of Fig.~\ref{fig:ACTlensing} we plot $M_\mathrm{5\,arcmin}$ as a function of tSZE-based halo mass estimate $M_{500}$ \citep{ACTDR5}. 
In producing the mean convergence maps, we account for the impact of contamination in the cluster sample. For this we use a catalog extracted along random lines of sight with the same selection cuts as those used for the real sample. After scaling the random catalog to the expected amount of contamination, we correct for the contribution of contaminants in each mass bin.

Correcting for the contaminating (i.e., non-cluster) systems requires accounting for the fact that 
the MCMF selection involves a redshift dependent richness cut.  This implies that each object in our cluster sample has an associated overdensity of passive galaxies, even in the rare cases where the object is associated with a tSZE noise fluctuation.  
Instead of assuming a value of zero we therefore use the average convergence around sources from the catalog of random lines of sight that share the same richness selection used for the real catalog. 
In fact we find a mean convergence for the sources identified along random lines of sight that is above zero with $2\sigma$ significance and that would correspond to a tSZE-based mass of $M_{500}\approx6\times10^{13}$ M$_\odot$.

Performing a detailed observable--mass scaling relation analysis is beyond the scope of this paper and would require a cluster optimized CMB lensing map \citep{SahaCMBlensing,SPT3Glensing}. Nevertheless, it is still interesting to perform a power law fit to the measurements to estimate their potential constraining power. In doing so we find that we can constrain the normalization of the scaling relation to a level of 6.7\%, and we find a slope of $1.15\pm0.15$. The slope is consistent with the expectation of unity between $M_{500}$ and $M_\mathrm{5\,arcmin}$ inlimit where the cluster is unresolved and $M_\mathrm{5\,arcmin}$ probing the total mass of the cluster.
We note that interpretation of the normalization of the scaling relation is not straight forward at this point due to the assumptions made on $M_\mathrm{5\,arcmin}$ as well as uncertainties on the scaling relation connecting tSZE-based mass estimate to true halo mass. 

Because the CMB lensing maps used here are not optimised for cluster lensing studies, the derived values need to be viewed with some caution. Imperfect modeling of the tSZE signal can bias the CMB lensing maps low, and the $\sim$5\,arcmin resolution of the map blurs the cluster signal somewhat.  Both effects reduce the signal-to-noise ratios provided here. A proper cluster focused CMB lensing analysis of the ACT-DR6 data set will therefore likely yield even better signal-to-noise than we present here.


\section{Conclusions}
In this work we present the ACT-DR5~MCMF cluster catalog. Using publicly released mm-wave maps and tSZE cluster finding code from the ACT collaboration \citep{ACTDR5maps,ACTDR5}, we create a cluster candidate list of 12,671 sources with SNR>4. A systematic follow-up of 97\% of all these ACT-DR5 candidates using LS-DR10 and WISE imaging data, allows us to exclude the noise fluctuations from the candidate list and create a cluster catalog with redshift and richness measurements. 

In contrast to the previously published ACT-DR5 catalog \citep{ACTDR5} we extract the ACT-D5 candidate list to lower SNR (SNR=3) than our subsequently imposed catalog threshold of SNR=4. This ensures a stable SNR-selected sample at SNR>4 and avoids additional selection effects that are apparent in the candidate list near the extraction threshold. 
Using MCMF derived richnesses of the candidates and along random lines of sight within the survey region, we characterize the ACT-DR5 candidate list and measure the contamination by noise fluctuations as a function of SNR threshold SNR$_\mathrm{min}$. We find excellent agreement between the measured contamination and that expected from pure Gaussian noise over almost four orders of magnitude in dynamic range of the candidate sample size. The tSZE candidate list has an initial contamination of 53\%.

We use the MCMF algorithm to construct a cluster catalog with controlled purity and completeness and define our baseline ACT-DR5~MCMF catalog of 6,237 clusters with a purity of $\sim$90\% and completeness with respect to the originial tSZE selection of 98\%. Our cluster catalog contains $\sim51$\% (2,080) more clusters than the previous ACT-DR5 catalog. The number of high-z ($z>1$) clusters exceeds 700, tripling the cluster sample at high redshifts compared to the previous catalog constructed with the same mm-wave data.

Overall, in comparison with the original ACT-DR5 cluster catalog \citep{ACTDR5}, the ACT-DR5~MCMF catalog provides 1) a true SNR-selected sample at SNR>4, 2) is $\sim51$\% larger overall and contains three times more z>1 clusters, 3) enables control over final catalog purity and 4) delivers higher completeness with respect to the tSZE selection. The last element is highlighted by the fact that our analysis leads to the confirmation of 327 new clusters at SNR>5. 
This catalog opens new opportunities for cosmological studies as well as the study of high redshift clusters and their components such as the ICM, galaxies and AGN. 

In addition to the baseline ACT-DR5~MCMF cluster catalog described above, we explain how subsets with much higher purity (lower contamination) can be selected to support, for example, a robust cosmological analysis.

We explore the properties of the ACT-DR5~MCMF cluster sample by comparing it to a deeper tSZE cluster sample from SPT and to the two largest ICM selected cluster catalogs currently available: the all-sky RASS-MCMF catalog and the eROSITA eRASS1 western Galactic hemisphere catalog.
Matching the ACT-DR5~MCMF catalog with the tSZE-based but deeper SPTpol~500d survey \citep{Bleem24} allows us to test and confirm the sample purity and completeness of the ACT-DR5~MCMF candidate list and cluster sample.
Over 1,000 ACT-DR5~MCMF clusters are matched to RASS-MCMF \citep{Klein23}, with the matched fraction rising rapidly to z$\sim$1 due to the differences in the X-ray and tSZE selections.  The matched fraction of RASS-MCMF sources within the ACT-DR5~MCMF catalog reaches 80-85\% at z>0.5, reflecting the expected contamination of the X-ray selected cluster sample.
Just over 1,300 ACT-DR5~MCMF clusters match to X-ray sources from eROSITA eRASS1 \citep{Merloni24}. We find that 43\% of the ACT-DR5~MCMF clusters with eRASS1 counterparts are classified as X-ray point sources rather than clusters and groups. This underscores the strong impact of the X-ray extent selection employed for deriving the eRASS1 group and cluster sample \citep{Bulbul24}. 
Our analysis underscores that within the context of the modest $\sim$0.5\,arcmin angular resolution of the eROSITA sky survey \citep{Predehl21}, an X-ray extent selected group and cluster catalog suffers from high incompleteness.

The average weak gravitational lensing convergence signal extracted from ACT-DR6 CMB lensing maps \citep{ACTDR6lensing} around ACT-DR5~MCMF clusters yields a detection significance of 16.4\,$\sigma$, the highest value reported so far for CMB lensing in a galaxy cluster sample. In addition, we find a 4.3\,$\sigma$ detection of CMB lensing for the $z>1$ ACT-DR5~MCMF cluster subsample. With a mean redshift of $z\approx1.2$, this subsample is currently the highest redshift cluster sample with a significant CMB lensing detection. As those significances are based on CMB lensing maps that have not been optimised for cluster lensing, the significances we present may be seen as conservative estimates.  We present the mean CMB lensing mass within 5\,arcmin ($M_\mathrm{5\,arcmin}$) to halo mass relation for the sample, where the halo masses are estimated using a previously calibrated tSZE observable-mass relation \citep{ACTDR5}. This exercise indicates that $M_\mathrm{5\,arcmin}$ scales as halo mass $M_{500}^{1.15\pm0.15}$ with the amplitude constrained with $\sim$6\% precision, underscoring the promise of the ACT-DR6 CMB lensing dataset for calibrating the masses of these ACT-DR5~MCMF clusters.

As shown in the case of ACT-CL~J1424.7+0642, an ACT-DR5 candidate with SNR=7.3 that is not confirmed by MCMF, we do expect to miss clusters-- especially at high redshift-- due to the limitations of the available follow-up data. Survey data from the operating Euclid mission \citep{EuclidMission} in near infrared will enable significant improvements in cluster followup in the next years.  Independent of that, the current ACT-DR5~MCMF catalog, including $\sim$700 clusters at z>1, is already an important step forward, opening a new large-sample window into the high redshift cluster population and the physical processes shaping the associated galaxy and AGN populations.

\begin{acknowledgements}
This work was initiated in part at the Aspen Center for Physics, which is supported by National Science Foundation grant PHY-2210452  Our thanks go out to Matt Hilton for supportive comments and the encouragement to undertake this project.  In addition, we thank the ACT Collaboration for their generosity in making their excellent dataset and toolkit available for the broader community. We acknowledge financial support from the MPG Faculty Fellowship program at MPE and from the Ludwig-Maximilians-Universit\"at (LMU Munich).
We gratefully acknowledge the NSF supported Legacy Survey program (see also https://www.legacysurvey.org/acknowledgment for detailed acknowledgements).
This research has made use of ``Aladin sky atlas'' developed at CDS, Strasbourg Observatory, France \citep{Aladin} and \textsc{topcat} \citep{Topcat}.
 \end{acknowledgements}

%
%
\bibliography{optid_refs}



\appendix
\section{Richness transformation factors}\label{app:richtrans}
Table~\ref{tab:oldnewrich} shows the first eleven entries of the transformation factor between $g,r,z,w1,w2$-based MCMF richness used in \cite{Klein23} and the updated $g,r,z,w1$-based richness presented in this work. The richness in \cite{Klein23} needs to be multiplied with the factor given in Table~\ref{tab:oldnewrich} to transform to the $g,r,z,w1$-based richness used for ACT-DR5~MCMF. The full table will be provided on CDS after acceptance of the paper.

\begin{table}
\caption{Transformation factors between $g,r,z,w1,w2$-based MCMF richness used in \cite{Klein23} and $g,r,z,w1$-based richness used in this work for a given redshift z. }\label{tab:oldnewrich}
\begin{tabular}{r r}
\hline
\hline
  \multicolumn{1}{c}{z} &
  \multicolumn{1}{c}{GRZW1W2\_TO\_GRZW1} \\
\hline
  0.0 & 0.8719359466775638\\
  0.001 & 0.8712321542454213\\
  0.002 & 0.8704275823591408\\
  0.003 & 0.8700663818489788\\
  0.004 & 0.8697660905334137\\
  0.005 & 0.8692880596168335\\
  0.006 & 0.8688464482133338\\
  0.007 & 0.8678339705471051\\
  0.008 & 0.8666583418494209\\
  0.009 & 0.865665771149266\\
  0.01 & 0.8649257435343979\\
  ... & ...\\
\hline\end{tabular} 
\end{table}

\section{Column description of the ACT-DR5~MCMF catalog}\label{app:ACTMCMFcols}
In Table~\ref{tab:columns} we describe the columns included on the released ACT-DR5~MCMF catalog. All entries related to tSZE-based measurements follow closely those provided in the original ACT-DR5 catalog presented \cite{ACTDR5} and we kindly refer the interested reader to that work for more detailed discussion of those entries. Entries related to redshift, richness and confirmation are provided from MCMF and correspond to the best result from combining LS DR-10 and unWISE MCMF runs.

\begin{table*}
\caption{Column names and description of the ACT-DR5~MCMF catalog}\label{tab:columns}
\begin{tabular}{l l}
\hline
\hline
  \multicolumn{1}{l}{Column name} &
  \multicolumn{1}{l}{Description} \\
 \hline
  NAME & Cluster name in the format ACT-CL JHHMM.m ± DDMM\\
  RADEG & R.A. in decimal degrees (J2000) of the SZ detection by ACT \\
  DECDEG & Decl. in decimal degrees (J2000) of the SZ detection by ACT\\
  SNR & Signal-to-noise ratio, optimized over all filter scales\\
  TEMPLATE & Name of the optimal matched filter template for this cluster\\
  TILENAME & Name of the ACT map tile in which the cluster was found \\
  FIXED\_SNR & Signal-to-noise ratio at the reference 2.4\,arcmin filter scale \\
  Y\_C & Central Comptonization parameter ($10^{-4}$) of the optimal matched filter template \\
  ERR\_Y\_C & Uncertainty on column Y\_C\\
 FIXED\_Y\_C & As Y\_C but using measured at the reference filter scale of 2.4\,arcmin\\
  FIXED\_ERR\_Y\_C & Uncertainty on column FIXED\_Y\_C   \\
  Z\_1\_COMB & Redshift of the best optical counterpart \\
  Z\_2\_COMB & Redshift of the second best counterpart  \\
  L\_1\_COMB & Richness of the best counterpart \\
  L\_2\_COMB & Richness of the second best counterpart\\
  EL\_1\_COMB & Uncertainty on column L\_1\_COMB \\
  EL\_2\_COMB & Uncertainty on column L\_2\_COMB \\
  EZ\_1\_COMB & Uncertainty on column Z\_1\_COMB \\
  EZ\_2\_COMB & Uncertainty on column Z\_2\_COMB \\
  F\_CONT\_1\_COMB & \fcont of best counter part \\
  F\_CONT\_2\_COMB & \fcont of second best counter part\\
  M500\_1\_COMB & tSZE-based estimate of $M_\mathrm{500c}$ in units of $10^{14} M_\odot$ using  Z\_1\_COMB  \\
  & (corresponds to M500c in \cite{ACTDR5})\\
  M500\_1\_COMB\_ERRPLUS & positive uncertainty to M500\_1\_COMB\\
  M500\_1\_COMB\_ERRMINUS & negative uncertainty to M500\_1\_COMB \\
  M500\_2\_COMB & As M500\_1\_COMB  but for  Z\_2\_COMB \\
  F\_CONT\_1\_COMB\_SNR45 & \fcont of best counter part for SNR>4.5 sample \\
  F\_CONT\_2\_COMB\_SNR45 & \fcont of second best counter part for SNR>4.5 sample\\
  M500\_1\_COMB\_CAL & Similar to M500\_1\_COMB  but rescaled with an calibration factor of $0.71\pm 0.07$ \\
  & (corresponds to M500\_cal in \cite{ACTDR5}) \\
  M500\_1\_COMB\_CAL\_ERRMINUS & positive uncertainty to M500\_1\_COMB\_CAL \\
  M500\_1\_COMB\_CAL\_ERRPLUS & negative uncertainty to M500\_1\_COMB\_CAL\\
  M500\_2\_COMB\_CAL & As M500\_1\_COMB\_CAL but using  Z\_2\_COMB\\
  Z\_SPEC\_1 & Spectroscopic redshift for the best optical counter part \\
  Z\_SPEC\_TYPE & Type of spec-z. 1: literature redshift, 2: >2 spec-z galaxies, 3: $\leq2$ spec-z galaxies\\
\hline\end{tabular}
\end{table*}


\end{document}